\documentclass[seceq,preprint]{ptptex}

\usepackage{graphicx}

\usepackage{amsmath}
\usepackage{amssymb}
\usepackage{mathrsfs}

\def\defscript{\mathscr}
\def\A{{\defscript A}}

\def\C{{\defscript C}}

\def\E{{\defscript E}}
\def\F{{\defscript F}}

\def\K{{\defscript K}}

\def\N{{\defscript N}}

\def\ifempty#1{\def\tmpdata{#1}\ifx\tmpdata\empty }

\def\linebreak{\hfill\break}

\def\bra<#1|{\langle #1\rvert}
\def\ket|#1>{\lvert#1 \rangle}
\def\braket<#1|#2>{\langle #1|#2 \rangle}

\def\mod{{\rm mod}}

\def\orth{\perp}
\def\pfrac#1#2{\left(\frac{#1}{#2}\right)}
\def\const{\text{const}}
\def\otop#1{\hbox{$#1\kern-0.1em$\llap{\hbox{\raise1.7ex\hbox{$\scriptstyle\circ$}}}} }

\def\inpare#1{\left(#1\right)}
\def\bigpare(#1){\left(#1\right)}
\def\inrbra#1{\left\{ #1 \right\}}
\def\insbra#1{\left[ #1 \right]}

\def\bigbra[#1]{\left[ #1 \right]}

\def\h{\hat }
\def\t{\tilde }

\def\b{\bar }

\def\tend{\rightarrow}

\def\therefore{\mbox{\setbox0=\hbox{X}\hbox{$\ldotp$}\raise0.7\ht0\hbox{$\ldotp$}\hbox{$\ldotp$}} \quad }
\def\because{\mbox{\setbox0=\hbox{X}\raise0.7\ht0\hbox{$\ldotp$}\hbox{$\ldotp$}\raise0.7\ht0\hbox{$\ldotp$}}\kern0pt }

\def\bm#1{\boldsymbol{#1}}

\def\RF{{{\mathbb R}}}

\def\maps{\rightarrow}

\def\upin{\hbox{\setbox0=\hbox{$\cup$} \vrule width 0.05 \wd0 height \ht0 depth 0pt \kern - 0.5\wd0 \box0 }}
\def\Frac(#1/#2){\left(\frac{#1}{#2}\right)}

\def\Im{{\rm Im\,}}
\def\Re{{\rm Re\,}}

\def\sdprod{\mathrel{{\setbox0=\hbox{$\displaystyle\times$}\lower0.3\wd0\hbox{$\stackrel{\box0}{\scriptstyle\sim}$}}}}

\def\tosigma#1,{%
    \ifx\tmpindex\relax \def\tmpindex{#1} \let\next=\tosigma
    \else \ifnum\tmpindex=0 1 \else \sigma_\tmpindex \fi
          \ifx#1\relax  \let\next=\relax
          \else \otimes \let\next=\tosigma \def\tmpindex{#1} \fi
    \fi \next}
\def\tspb(#1){\let\tmpindex=\relax\tosigma#1,\relax,}
\def\pd{\partial}

\def\Lie{\hbox{\rlap{$\cal L$}$-$}}

\def\AdS{{\rm AdS}}

\def\THB{{\mathbb T}}
\def\VHB{{\mathbb V}}
\def\SHB{{\mathbb S}}

\def\Eq#1{\begin{equation} #1 \end{equation}}
\def\Eqn#1{\Eq{#1 \nonumber}}
\def\Eqr#1{\begin{eqnarray} #1 \end{eqnarray}}
\def\Eqrn#1{\begin{eqnarray*} #1 \end{eqnarray*}}
\def\Eqrsub#1{\begin{subequations}\Eqr{#1}\end{subequations}}
\def\Eqrsubl#1#2{\begin{subequations}
  \expandafter\ifx\csname Rlabel\endcsname \relax \label{#1}
  \else \Rlabel{#1} \fi \Eqr{#2}\end{subequations}}
\def\Bitm{\begin{itemize}}
\def\Eitm{\end{itemize}}
\def\Blist#1#2{\begin{list}{#1}{\parsep=0pt \itemsep=0pt%
  \listparindent=0pt #2}}
\def\Elist{\end{list}}
\long\def\ignore#1#2{\def\ignoreflag{#1}\long\def\tmptext{#2}
  \ifnum\ignoreflag>1 #2 \fi}

\def\FigDir{./}

\def\THB{{\mathbb T}}
\def\VHB{{\mathbb V}}
\def\SHB{{\mathbb S}}

\def\paragraph#1{\par\smallskip\noindent{\bf #1}\ }
\notypesetlogo

\begin{document}

\preprintnumber[3cm]{KEK-Cosmo-2}

\title{
Perturbations and Stability of\\ Higher-Dimensional Black 
Holes\footnote{Based on the lecture given at the 4th Aegean Summer School on Black Holes, 17-22 September 2007. A slightly condensed version will be published as a part of the lecture note collection from Springer.}
}

\author{
Hideo Kodama
\footnote{E-mail: Hideo.Kodama@kek.jp}
}
\inst{
Cosmophysics Group, IPNS, KEK and 
the Graduate University of Advanced Studies,
1-1 Oho, Tsukuba 305-0801, Japan
}


\abst{
 In this lecture, I explain the gauge-invariant formulation for perturbations of background spacetimes with untwisted homologous Einstein fibres, which include lots of practically important spacetimes such as static black holes, static black branes and rotating black holes in various dimensions. As applications, we discuss the stability of static black holes in higher dimensions and flat black branes.
}

\maketitle

\section{Introduction}

Perturbation analysis is a very powerful tool to investigate the dynamical response of a system against small disturbances. In particular, in general relativity whose fundamental equations are quite hard to solve analytically in general due to their nonlinearity and strong couplings, perturbation analysis of exact solutions plays crucial roles in physical and astrophysical problems. The most successful example is the perturbative studies of cosmological perturbations, which has in particular provided the foundation for the present structure formation theory and the precise observational cosmology in terms of CMB and gravitational waves.

Another important example is the perturbative studies of black holes. Such an investigation was first systematically done for the Schwarzschild black hole by Regge and Wheeler\cite{Regge.T&Wheeler1957} in 1957. In particular, they succeeded in reducing the Einstein equations for odd-parity perturbations, which is called vector perturbations in the present lecture, to a single master ODE, which is called the Regge-Wheeler equation now. The formulation was extended to even-parity perturbations (scalar perturbations in this lecture) by Zerilli thirteen years later\cite{Zerilli.F1970}, and the master equation called the Zerilli equation was derived for such perturbations. Soon later, Teukolsky succeeded in deriving similar Master equations for perturbations of the Kerr black hole\cite{Teukolsky.S1972}.

The original purpose of these formulations appears to have been to study gravitational emissions from particles plunging into or orbiting around black holes. However, it was soon recognised\cite{Vishveshwara.C1970,Price.R1972} that the formulation can be used to study the stability of black holes, which is also a practically important problem in determining the final fate of gravitational collapse. Actually, the asymptotically flat neutral and charged black holes were shown to be stable (for the proof and  its historical background, see the excellent book by Chandrasekhar\cite{Chandrasekhar.S1983B}). This together with the uniqueness theorem for black holes in the asymptotically flat electrovac system\cite{Heusler.M1996B} now provide the basis of the current black hole astrophysics.

These results on four-dimensional black holes are practically sufficient for investigations of low energy phenomena. However, taking account of the higher-dimensionality of the present candidates for the unified theory, it is likely that higher-dimensional black holes are formed in the early universe and in extremely high energy astrophysical phenomena as well as in particle accelerators. In fact, motivated by this expectation, lots of work has been done on higher-dimensional black holes in various theories, and astonishing discoveries have been obtained. In particular, it is now widely recognised that black hole uniqueness does not hold in higher dimensions, except for static black holes\cite{Emparan.R2004,Kodama.H2004a}. Furthermore, a full list of regular black holes has not been obtained even in five dimensions\cite{Kol.B2006,Elvang.H&Emparan&Figueras2007}(Cf. \citen{Hollands.S&Yazadjiev2007A,Morisawa.Y&Tomizawa&Yasui2007A}). 

In this situation, perturbative analysis of exact solutions found so far is expected to be quite useful for the study of stability and uniqueness of higher-dimensional black holes\cite{Kodama.H2004}. In particular, it will be a great help if we can reduce the Einstein equations for perturbations of higher dimensional black holes to decoupled master equations, as in the four-dimensional case. 
In this lecture, we show that we can really reduce the perturbation equations to decoupled mater equations for some classes of black holes and study the stability with the help of them. We also point out that such a reduction is not always possible.

The remaining part is organized as follows. First, in the next section, we briefly overview the present status of the black hole stability issue in four and higher dimensions. Then, in \S\ref{kod:sec:GIPT}, we explain the basic aspects of the gauge-invariant formulation for perturbations of a general class of background spacetimes that can be written as a warped product of a lower dimensional spacetime and an Einstein space. In \S\ref{kod:sec:SBH}, we apply this formulation to static black holes and discuss their stability. Next, in \S\ref{kod:sec:FBB}, we study the stability of flat black branes and point out the non-hermitian nature of the perturbation equations of this system. Section \ref{kod:sec:Summary} is devoted to brief summary and discussion.

\section{Present Status of the Black Hole Stability Issue}

In this section, we briefly overview the present status of the investigations of the stability problem of black holes. It is far from complete.

\subsection{Four Dimensions}

The present status of the stability issue for four-dimensional black holes is summarised as follows:

\Bitm
\item Stable
\Bitm 
\item Schwarzschild black hole\cite{Vishveshwara.C1970,Price.R1972,Wald.R1979,Wald.R1980}
\item Reissner-N{\"o}rdstrom black hole\cite{Chandrasekhar.S1983B}
\item AdS/dS (charged) black holes\cite{Ishibashi.A&Kodama2003,Kodama.H&Ishibashi2004}
\item Kerr black hole\cite{Whiting.B1989}
\item Skyrme black hole (non-unique system)\cite{Heusler.M&Droz&Straumann1991,Heusler.M&Droz&Straumann1992,Heusler.M&Straumann&Zhou1993}
\Eitm
\item Unstable
\Bitm
\item YM black hole (non-unique system)\cite{Straumann.N&Zhou1990,Zhou.Z&Straumann1991}
\item Kerr-AdS black hole ($\ell\Omega_h<1, r_h\ll \ell$)\cite{Cardoso.V&Dias&Yoshida2006}.
\Eitm
\Eitm

As is seen from this list, the stability is established for all AF black holes with connected horizon in the Einstein-Maxwell system, except for the charged rotating black hole (Kerr-Newman black hole). This is because the perturbation equations have not been reduced to decoupled single master equations for this system.  

In the asymptotically adS/dS case, the stability of static black holes have been established with the help of mater equations derived by Cardoso and Lemos\cite{Cardoso.V&Lemos2001}. In contrast, in the rotating case, it was conjectured that large Kerr-AdS black holes are stable, while small ones are superradiant unstable\cite{Hawking.S&Reall1999,Cardoso.V&Dias2004}. Recently, the conjecture was shown to be true in the limit of slow rotation and small horizon.

\subsection{Higher Dimensions}

In contrast to the four-dimensional case, in addition to conventional black holes, there exist different kinds of black objects such as black strings, black branes, Kaluza-Klein black holes, Kaluza-Klein bubbles and black tubes  in higher dimensions. The classification of these black objects  is far from complete, and rather a little is know about the stability of known solutions. For example, concerning the asymptotically flat/dS/adS black holes and black branes, the present status of the stability issue is summarised as follows:
\Bitm
\item Stable
\Bitm
\item AF vacuum static (Schwarzschild-Tangherlini)\cite{Ishibashi.A&Kodama2003}
\item AF charged static ($D=5,6-11$)\cite{Kodama.H&Ishibashi2004,Konoplya.R&Zhidenko2007}
\item dS vacuum static ($D=5,6,7-11$), dS charged static ($D=5,6-11$)\cite{Ishibashi.A&Kodama2003,Kodama.H&Ishibashi2004,Konoplya.R&Zhidenko2007}
\item BPS charged black branes (in type II SUGRA)\cite{Gregory.R&Laflamme1994,Hirayama.T&Kang&Lee2003}
\Eitm
\item Unstable
\Bitm
\item AF/adS static black string and AF black branes (non-BPS)\cite{Gregory.R&Laflamme1993,Gregory.R&Laflamme1995,Gregory.R2000,Hirayama.T&Kang2001,Hirayama.T&Kang&Lee2003,Kang.G2004,Seahra.S&Clarkson&Maartens2005,Kudoh.H2006,Brihaye.Y&Delsate&Radu2007A}
\item Rapidly rotating special Kerr-AdS black holes\cite{Kunduri.H&Lucietti&Reall2006}
\Eitm
\Eitm

In this list, the stability of static black holes in higher dimensions ($D>4$) has been proved analytically only for $D=5$ in the AF/dS charged case and for $D=5,6$ in the dS neutral case, as explained in \S\ref{kod:sec:SBH}. The stability in other dimensions up to $D=11$ for these black holes was proved numerically\cite{Konoplya.R&Zhidenko2007}. It is expected that the same result holds for $D>11$ as well. 

In contrast, in the asymptotically adS case, stability in $D>4$ is not certain even for neutral static black holes. This is a delicate problem because instability is expected for rotating adS black holes\cite{Hawking.S&Reall1999,Cardoso.V&&2004,Cardoso.V&Yoshida2005,Cardoso.V&Dias&Yoshida2006,Kodama.H2007A}(Cf. Ref. \citen{Carter.BMN&Neupane2005}). This instability is understood to arise from the combination of superradiance due to a rotating black hole and the time-like nature of the adS infinity. Some people conjectured that this superradiance also invokes instabilities in doubly spinning black rings\cite{Dias.O2006} and Kerr black branes of the form Kerr$_4\times\RF^p$ \cite{Cardoso.V&Yoshida2005}.

The most impressive result about the stability of black objects in higher dimensions is the discovery of the Gregory-Laflamme instability of black strings/branes\cite{Gregory.R&Laflamme1993}. Since then, a large amount of work has been done on the classification of black holes and black strings/branes in the $S^1$/torus compactified system and  their stability. These researches revealed a rich structure of the phase diagram for such systems as well as new instabilities (for review, see Refs. \citen{Kol.B2006,Harmark.T&Niarchos&Obers2007A}). However, no clear understanding has been obtained about the origin and fate of these instabilities.  It is partly because most of the researches were done by numerical methods. In \S\ref{kod:sec:FBB}, we point out some features that may be obstacles against the analytic approach.

Finally, we have to emphasize that very little is known about the stability of asymptotically flat solutions with rotating black objects. For example, rapidly rotating Myers-Perry solutions were conjectured to suffer from a Gregory-Laflamme type instability\cite{Emparan.R&Myers2003}, but it has not been proved by any exact analysis. Black rings are also expected to be unstable because of their similarity to black string solutions, but no exact proof has been presented.

\section{Gauge-invariant Perturbation Theory}
\label{kod:sec:GIPT}

In this section, we explain the basic idea and techniques of the gauge-invariant formulation of perturbations\cite{Bardeen.J1980,Kodama.H&Sasaki1984} for a class of background spacetimes that includes static black hole spacetimes as special case. 

\subsection{Background Solution}

\subsubsection{Ansatz}

We assume that a background spacetime can be locally written as the warped product of a $m$-dimensional spacetime $\N$ and an $n$-dimensional Einstein space $\K$ as
\Eq{
M^{n+m} \approx \N\times \K \ni (z^M)=(y^a,x^i)
}
and has the metric
\Eq{
ds^2= g_{MN}dz^M dz^N
  = g_{ab}(y)dy^a dy^b + r(y)^2 d\sigma_n^2,
\label{kod:Background:metric}
}
where $d\sigma_n^2=\gamma_{ij}dx^idx^j$ is an $n$-dimensional Einstein space $\K$ satisfying the condition
\Eq{
\hat R_{ij}= (n-1)K \gamma_{ij}.
}
Note that for $n\le 3$ the Einstein space is automatically a constant curvature space, while for $n>3$, $\K$ does not have a constant curvature generically.

For this type of spacetimes, we can express the covariant derivative $\nabla_M$, the connection coefficients $\Gamma^M_{NL}(z)$ and the curvature tensor $R_{MNLS}(z)$ in terms of the corresponding quantities for $\N^m$ and $\K^n$. We denote them $D_a,{}^m\!\Gamma^a_{bc}(y), {}^m\! R_{abcd}(y)$ and 
$\hat D_i, \hat\Gamma^i_{jk}(x), \hat R_{ijkl}(x)$ respectively. In particular, the curvature tensor can be expressed as
\Eq{
R^a{}_{bcd}={}^m\! R^a{}_{bcd},\quad
R^i{}_{ajb}=-\frac{D_aD_b r}{r}\delta^i_j,\quad
R^i{}_{jkl}={}^m\!R_{abcd}-(Dr)^2(\delta^i_k\gamma_{jl}-\delta^i_l\gamma_{jk}).
}
Hence,the non-vanishing components of the Einstein tensor are given by 
\Eqrsub{
\hspace*{-1cm} G_{ab} &=& {}^m\!G_{ab}-\frac{n}{r}D_aD_b r 
-\left[\frac{n(n-1)}{2}\frac{K-(Dr)^2}{r^2}
-\frac{n}{r}\square r\right]g_{ab} \\
\hspace*{-1cm} G^i_j &=& \left[-\frac{1}{2}{}^m\!R-\frac{(n-1)(n-2)}{2}
\frac{K-(Dr)^2}{r^2}+\frac{n-1}{r}\square r \right]\delta^i_j.
}
From this and the Einstein equations $ G_{MN}+{\Lambda}  g_{MN}={\kappa}^2  T_{MN}$, it follows that the energy-momentum tensor of the background solution should take the form
\Eq{
{ T}_{ab}={ T}_{ab}(y),\ 
{ T}_{ai}=0,\ 
{ T}^i_j={ P(y)}\delta^i_j.
}
%

\subsubsection{Examples}

This class of background spacetimes include quite a large variety of important solutions to the Einstein equations in four and higher dimensions. 

\Bitm
\item[1.] {\bf Robertson-Walker universe}: $m=1$ and $\K$ is a constant curvature space.
\Eqn{
ds^2= -dt^2 + a(t)^2 d\sigma_n^2.
}
The gauge-invariant formulation was first introduced for perturbations of this background by Bardeen\cite{Bardeen.J1980} and applied to realistic cosmological models by the author\cite{Kodama.H1984,Kodama.H1985,Kodama.H&Sasaki1984}.
\item[2.] {\bf Braneworld model}: $m=2$ (and $\K$ is a constant curvature space). For example, the metric of $\AdS^{n+2}$ spacetime can be written
\Eq{
ds^2=\frac{dr^2}{1-\lambda r^2} -(1-\lambda r^2) dt^2  + r^2 d\Omega_n^2.
}
The gauge-invariant formulation of this background was first discussed by Mukohyama\cite{Mukohyama.S2000} and then applied to the braneworld model taking account of the junction conditions by the author and collaborators\cite{Kodama.H&Ishibashi&Seto2000}.
\item[3.] {\bf Higher-dimensional static Einstein black holes}: $m=2$ and $\K$ is a compact Einstein space. For example, for the Schwarzschild-Tangherlini black hole, $\K=S^n$. In general, the generalised Birkhoff theorem says\cite{Kodama.H&Ishibashi2004} that the electrovac solutions of the form \eqref{kod:Background:metric} with $m=2$ to the Einstein equations are exhausted by the Nariai-type solutions such that $M$ is the direct product of a two-dimensional constant curvature spacetime $\N$ and an Einstein space $\K$ with $r=\const$ and the black hole type solution whose metric is given by
\Eqr{
&& ds^2=  \frac{dr^2}{f(r)}- f(r) dt^2 +r^2 d\sigma_n^2;\\
&& f(r)= K- \frac{2M}{r^{n-1}} + \frac{Q^2}{r^{2n-2}} -\lambda r^2.
\label{kod:background:sbh}}
The gauge-invariant formulation for perturbations was applied to this background to discuss the stability of static black holes by the author and collaborators\cite{Kodama.H&Ishibashi2003,Ishibashi.A&Kodama2003,Kodama.H&Ishibashi2004}. This application is explained in the next section. 
\item[4.] {\bf Black branes}: $m=2+k$ and $\K=\text{Einstein space}$. In this case, the spacetime factor $\N$ is the product of a two-dimensional black hole sector and a $k$-dimensional brane sector:
\Eq{
ds^2 = \frac{dr^2}{f(r)} - f(r)dt^2 + d\bm{z}\cdot d\bm{z}
  + r^2 d\sigma_n^2.
}
One can also generalise this background to introducing a warp factor in front of the black hole metric part. The stability of this background for the case in which $\K$ is an Euclidean space is discussed in \S\ref{kod:sec:FBB}.
\item[5.] {\bf Higher-dimensional rotating black hole} (a special Myers-Perry solution): $m=4$ and $\K=S^n$.
\Eq{
ds^2 =  g_{rr}dr^2 + g_{\theta\theta}d\theta^2
       +g_{tt}dt^2+2g_{t\phi}dtd\phi+g_{\phi\phi}d\phi^2
       + r^2\cos^2\theta d\Omega_n^2,
}
where all the metric coefficients are functions only of $r$ and $\theta$. The stability of this background was studied in Ref. \citen{Kodama.H2007A}.
\item[6.] {\bf Axisymmetric spacetime}: $m$ is general and $n=1$.
\Eitm

\subsection{Perturbations}

\subsubsection{Perturbation equations}

In order to describe the spacetime structure and matter configuration $(\tilde M, \tilde g,\tilde\Phi)$ as a perturbation from a fixed background $(M,g,\Phi)$, we introduce a mapping
\Eq{
F: \text{background\ }M \maps \tilde M,
}
and define perturbation variables on the fixed background spacetime as follows:
\Eqrsub{
&& h:=\delta g= F^* \tilde g - g,\\
&& \phi:=\delta \Phi= F^* \tilde \Phi -\Phi.
}
Then, if the perturbation variables have small amplitudes, the Einstein equations and the other equations for matter can be described by linearised equations well. For example, in terms of the variable
\Eq{
\psi_{\mu\nu}=h_{\mu\nu}-\frac{1}{2}hg_{\mu\nu},
}
the linearised Einstein equations can be written as
\Eqr{
&& \triangle_L \psi_{\mu\nu}
   +\nabla_\mu\nabla_\alpha\psi^\alpha_\nu
   +\nabla_\nu\nabla_\alpha\psi^\alpha_\mu
   -\nabla^\alpha\nabla^\beta \psi_{\alpha\beta}g_{\mu\nu}
   +R^{\alpha\beta}\psi_{\alpha\beta}g_{\mu\nu}
   -R\psi_{\mu\nu}
\nonumber\\
&& = 2\kappa^2\delta T_{\mu\nu}.
}
where $\triangle_L$ is the Lichnerowicz operator defined by
\Eq{
\triangle_L \psi_{\mu\nu}:=-\Box\psi_{\mu\nu}
 +R_{\mu\alpha}\psi^\alpha_\nu + R_{\nu\alpha}\psi^\alpha_\mu 
   -2R_{\mu\alpha\nu\beta}\psi^{\alpha\beta}.
}

\subsubsection{Gauge problem}

\begin{figure}[t]
\centerline{
\includegraphics*[width=10cm]{\FigDir/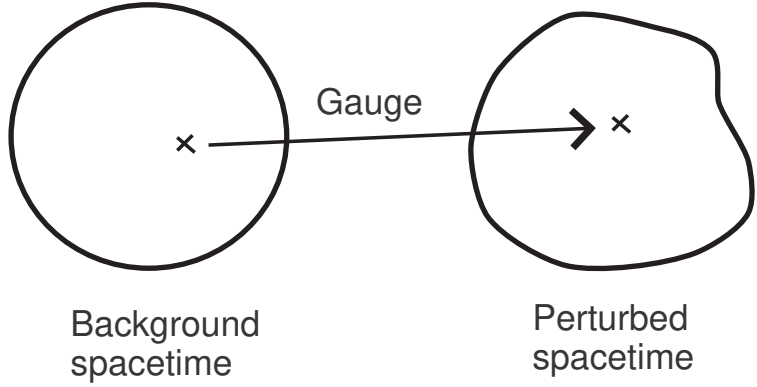}
}
\caption{Gauge transformation}
\end{figure}


For a different mapping $F'$, these perturbation variables defined above change their values, which has no physical meaning and can be regarded as a kind of gauge freedom. Because $F$ and $F'$ are related by a diffeomorphism, the corresponding changes of the variables are identical to the transformation of the variables with respect to the transformation $f=F'{}^{-1}F$. In the framework of linear perturbation theory, we can restrict considerations to infinitesimal changes of $F$. Hence, $f$ is expressed in terms of an infinitesimal transformation $\xi^\mu$ as
\Eq{
\bar\delta x^\mu= x^\mu(f(p)) -x^\mu(p)=\xi^\mu,
}
and the gauge transformations are expressed as
\Eqrsub{
&& \bar\delta h_{\mu\nu}= -\Lie_\xi g_{\mu\nu} \equiv -\nabla_\mu \xi_\nu - \nabla_\nu \xi_\mu,\\
&& \bar\delta \phi=-\Lie_\xi \Phi.
}
From its origin, the perturbation equations including the linearised Einstein equations given above are invariant under this gauge transformation.

To be specific, for our background spacetime, the metric perturbation transforms as
\Eqrsub{
&& \bar\delta h_{ab}=-D_a\xi_b-D_b\xi_a, 
\\
&& \bar\delta h_{ai}=-r^2D_a\left(\frac{\xi_i}{r^2}\right)-\hat D_i\xi_a,
\\
&& \bar\delta h_{ij}=-\hat D_i\xi_j-\hat D_j\xi_i-2rD^ar \xi_a \gamma_{ij}
}
and the perturbation of the energy-momentum tensor $\tau_{\mu\nu}=\delta T_{\mu\nu}$ transforms as
\Eqrsub{
&& \bar\delta\tau_{ab}=-\xi^cD_c{T}_{ab}
      -{T}_{ac}D_b\xi^c-{T}_{bc}D_a\xi^c,
\\
&& \bar\delta\tau_{ai}=-{T}_{ab}\hat D_i\xi^b
      -r^2{P}D_a(r^{-2}\xi_i),
\\
&& \bar\delta\tau_{ij}=-\xi^aD_a(r^2{P})\gamma_{ij}
       -{P}(\hat D_i\xi_j+\hat D_j\xi_i)
}

In order to remove this gauge freedom, one of the following two approaches is adopted in general:
\Bitm
\item[i)] {\bf Gauge fixing method}: this method is direct, but it is rather difficult to find relations between perturbation variables in different gauges in general.
\item[ii)] {\bf Gauge-invariant method}: this method describe the theory only in terms of gauge-invariant quantities. Such quantities have non-local expressions in terms of the original perturbation variables in general.
\Eitm
These two approaches are mathematically equivalent, and a gauge-invariant variable can be regarded as some perturbation variable in some special gauge in general. Therefore, the non-locality of the gauge-invariant variables implies that the relation of two different gauges are non-local.

\subsubsection{Tensorial decomposition of perturbations}

In this lecture, we focus on the gauge-invariant approach to perturbations and explain that in the class of background spacetimes described above, we can locally construct fundamental gauge invariant variables with help of harmonic expansions. This construction becomes more transparent if we decompose the perturbation variables into components of specific tensorial types. This decomposition also helps us to divide the coupled set of perturbation equations into decoupled smaller subsets, and in some cases into single master equations.

First of all,  note that the basic perturbation variables $h_{MN}$ and $\tau_{MN}$ can be classified into the following three algebraic types according to their transformation property as tensors on the $n$-dimensional space $\K$:
\Bitm
\item[i)] Spatial scalar: $h_{ab}, \tau_{ab}$
\item[ii)] Spatial vector: $h_{ai}, \tau^a_i$
\item[iii)] Spatial tensor: $h_{ij}, \tau^i_j$
\Eitm
%

Among these, spatial vectors and tensors can be further decomposed into more basic quantities. First, we decompose a vector field $v_i$ on $\K$ into a scalar field $v^{(\rm{s})}$ and a transverse vector $v^{(\rm{t})}_i$ as
\Eq{
v_i=\hat D_i v^{(\rm{s})} + v^{(\rm{t})}_i;\quad
\hat D_i v^{(\rm{t})i}=0. 
}
Then, from the relation
\Eq{
\hat \triangle v^{(\rm{s})}=\hat D_i v^i,
}
the component fields $v^{(\rm{s})}$ and $v^{(\rm{t})}_i$ can be uniquely determined from $v_i$ up to the ineffective freedom in $v^{(\rm{s})}$ to add a constant, provided that this Poisson equation has a unique solution on $\K$ up to the same freedom. For example, when $\K$ is compact and closed, this condition is satisfied. 

Next, we decompose a symmetric tensor field of rank 2 on $\K$ as
\Eqrsub{
&& t_{ij}=\frac{1}{n}t g_{ij}+ \hat D_i\hat D_j s
        -\frac{1}{n}\hat \triangle s g_{ij}
        + \hat D_i t_j +\hat D_j t_i 
        + t^{(\rm{tt})}_{ij}; 
\\
&& \hat D_i t^i=0,\quad 
   t^{(\rm{tt})i}_i=0,\quad \hat D_i t^{(\rm{tt})i}_j=0.
}
Here, $t$ is uniquely determined as $t=t^i_i$. Further, from the relations derived from this definition, 
\Eqrsub{
&& \hat \triangle(\hat\triangle +nK) s
    =\frac{n}{n-1}\left(\hat D_i\hat D_jt^{ij}
     -\frac{1}{n}\hat\triangle t\right),
\\
&& [\hat\triangle+(n-1)K]t^i=(\delta^i_j-\hat D^i\hat\triangle^{-1}\hat D_j)
   (\hat D_mt^{jm}-n^{-1}\hat D^jt),
}
$s$ and $t_i$, hence $t^{(\rm{tt})}_{ij}$, can be uniquely determined from $t_{ij}$ up to the addition of ineffective zero modes, provided that these Poisson equations have solutions unique up to the same ineffective freedom.

After these decompositions of vectors and tensors to basic components, we can classify these components into the following three types:
\Bitm
\item[i)] Scalar type: 
  $v^i=\hat D^i v^{(\rm{s})},\ 
  t_{ij}=\frac{1}{n}t g_{ij}+ \hat D_i\hat D_j s
         -\frac{1}{n}\hat\triangle s g_{ij}$.
\item[ii)] Vector type: 
  $v_i=v^{(\rm{t})}_i,\ 
  t_{ij}= \hat D_i t_j +\hat D_j t_i$.
\item[iii)] Tensor type: 
  $v^i=0,\ t^i_j=t^{(\rm{tt})^i}_j$.
\Eitm
We call these types {\em reduced tensorial types}. 
In the linearised Einstein equations, through the covariant differentiation and tensor-algebraic operations, quantities of different algebraic tensorial types can appear in each equation. However, in the case in which $\K$ is a constant curvature space, perturbation variables belonging to different reduced tensorial types do not couple in the linearised Einstein equations if we decompose these perturbation equations into reduced tensorial types as well, because there exists no quantity of the vector or the tensor type in the background except for the metric tensor. The same result holds even in the case in which $\K$ is an Einstein space with non-constant curvature, because the only non-trivial background tensor other than the metric is the Weyl tensor that can only transform a 2nd rank tensor to a 2nd rank tensor.

Here, note that gauge transformations can be also decomposed into reduced tensorial types, and the gauge transformation of each type affects only the decomposed perturbation variables of the same reduced tensorial type. Hence, gauge-invariant variables can be constructed in each reduced tensorial types independently.

\subsection{Tensor Perturbation}

Let us start from the tensor-type perturbation, for which the argument is simplest.

\subsubsection{Tensor Harmonics}

We utilise tensor harmonics to expand tensor-type perturbations. They are defined as the basis for 2nd-rank symmetric tensor fields satisfying the following eigen-value problem:
\Eq{
(\hat\triangle_L -\lambda_L)\THB_{ij}=0;\quad
\THB^i_i=0, \quad \hat D_j \THB^j_i=0,
}
where $\hat\triangle_L$ is the Lichnerowicz operator on $\K$ defined by
\Eq{
\hat\triangle_L h_{ij}:= -\h D\cdot \h D h_{ij} - 2 \h R_{ikjl} h^{kl} 
 + 2(n-1) K h_{ij}.
}
When $\K$ is a constant curvature space, this operator is related to the Laplace-Beltrami operator by
\Eq{
\h\triangle_L = -\h\triangle + 2nK,
}
and, $\THB_{ij}$ satisfies
\Eq{
(\h\triangle + k^2) \THB_{ij}=0;\quad
  k^2=\lambda_L - 2nK.
}
We use $k^2$ in the meaning of $\lambda_L-2nK$ from now on when $\K$ is an Einstein space with non-constant sectional curvature.

The harmonic tensor has the following basic properties:
\Bitm
\item[1.] {\bf Identities}: Let $T_{ij}$ be a symmetric tensor of rank 2 satisfying
\Eqn{
T^i_i=0,\quad
D^j T_{ij}=0.
}
Then, the following identities hold:
\Eqrn{
&& 2D_{[i}T_{j]k} D^{[i}T^{j]k}= 2D^i(T_{jk}D^{[i}T^{j]k})
   +T_{jk}\insbra{-\triangle T^{jk} + R^j_l T^{lk}
     +R_i{}^{jk}{}_l T^{il}},\\
&& 2D_{(i}T_{j)k} D^{(i}T^{j)k}= 2D^i(T_{jk}D^{(i}T^{j)k})
   +T_{jk}\insbra{-\triangle T^{jk} - R^j_l T^{lk}
     -R_i{}^{jk}{}_l T^{il}}.
}
On the constant curvature space with sectional curvature $K$, these identities read
\Eqrn{
&& 2D_{[i}T_{j]k} D^{[i}T^{j]k}= 2D^i(T_{jk}D^{[i}T^{j]k})
   +T_{jk}( -\triangle +nK) T^{jk},\\
&& 2D_{(i}T_{j)k} D^{(i}T^{j)k}= 2D^i(T_{jk}D^{(i}T^{j)k})
   +T_{jk}( -\triangle -nK) T^{jk}.
}
\item[2.] {\bf Spectrum}: When $\K$ is a compact and closed space with constant sectional curvature $K$, these identities lead to the following condition on the spectrum of $k^2$:
\Eq{
 k^2\ge n|K|.
}
In contrast, when $\K$ is not a constant curvature space, no general lower bound on the spectrum $k^2$ is known.
\item[3.] When $\K$ is a two-dimensional surface with a constant curvature $K$, a symmetric 2nd rank harmonic tensor that is regular everywhere can exist only for$K\le0$: for $T^2$($K=0$), the corresponding harmonic tensor $T_{ij}$ becomes a constant tensor in the coordinate system such that the metric is written $ds^2=dx^2+dy^2$($k^2=0$); for $H^2/\Gamma$($K=-1$), a harmonic tensor corresponds to an infinitesimal deformation of the moduli parameters.
\item[4.] For $\K=S^n$, the spectrum of $k^2$ is given by
\Eq{
k^2= l(l+n-1) -2;\quad l=2,3,\cdots,
}
\Eitm

\subsubsection{Perturbation equations}

The metric and energy-momentum perturbations can be expanded in terms of the tensor harmonics as
\Eq{
h_{ab}=0,\quad 
h_{ai}=0,\quad
h_{ij}=2r^2 H_T \THB_{ij},
}
\Eq{
\tau_{ab}=0,\ 
\tau^a_{i}=0,\ 
\tau^i_{j}=\tau_T \THB^i_{j}.
}
%
Since the coordinate transformations contains no tensor-type component, $H_T$ and $\tau_T$ are gauge invariant by themselves:
\Eqrsub{
&& \xi^M=\bar\delta z^M=0;\\
&& \bar\delta H_T=0, \ 
   \bar\delta \tau_T=0.
}
%

Only the $(i,j)$-component of the Einstein equations has the tensor-type component:
\Eq{
-\square H_T-\frac{n}{r}Dr\cdot DH_T+ 
\frac{k^2+2K}{r^2}H_T={\bar\kappa}^2\tau_T. 
\label{kod:BulkPerturbationEq:tensor}
}
Here, $\square=D^aD_a$ is the D'Alembertian in the $m$-dimensional spacetime $\N$. Thus, the Einstein equations for tensor-type perturbations can be always reduced to the single master equation on our background spacetime.

\subsection{Vector Perturbation}

\subsubsection{Vector harmonics}

We expand transverse vector fields in terms of the complete set of harmonic vectors defined by the eigenvalue problem
%
\Eq{
(\hat\triangle +k^2)\VHB_i=0;\quad
 \hat D_i \VHB^i=0.
}
Tensor fields of the vector-type can be expanded in terms of the harmonic tensors derived from these vector harmonics as
\Eq{
\VHB_{ij}=-\frac{1}{2k}(\hat D_i\VHB_j+\hat D_j\VHB_i).
}
They satisfy
\Eqrsub{
&& \left[\hat \triangle +k^2-(n+1)K\right]\VHB_{ij}=0,\\
&& \VHB^i_i=0,\quad
   \hat D_j \VHB^j_i=\frac{k^2-(n-1)K}{2k}\VHB_i.
}

Here, there is one subtle point; $\VHB_{ij}$ vanishes when $\VHB_i$ is a Killing vector. For this mode, from the above relations, we have
\Eq{
k^2=(n-1)K.
}
We will see below that the converse holds when $\K$ is compact and closed. We call these modes {\em exceptional modes}.

Now, we list up some basic properties of the vector harmonic relevant to the subsequent discussions.
\Bitm
\item[1.] {\bf Spectrum}: In an $n$-dimensional Einstein space $\K$ satisfying
\Eq{
R_{ij}=(n-1)K g_{ij},
}
we have
\Eqrsub{
&& 2D_{[i} V_{j]}D^{[i}V^{j]}=2D_i(V_jD^{[i}V^{j]})
   +V_j\insbra{-\triangle +(n-1)K} V^j,\\
&& 2D_{(i} V_{j)}D^{(i}V^{j)}=2D_i(V_jD^{(i}V^{j)})
   +V_j\insbra{-\triangle -(n-1)K} V^j.
}
When $\K$ is compact and closed, from the integration of these over $\K$, we obtain the following general restriction on the spectrum of $k^2$:
\Eq{
k^2\ge (n-1)|K|.
}
Here, when the equality holds, the corresponding harmonic vector becomes a Killing vector for $K\ge0$ and a harmonic 1-form for $K\le0$, respectively. 
\item[2.] For$\K^n=S^n$, we have
\Eq{
k^2=\ell(\ell+n-1)-1,\quad(\ell=1,2,\cdots).
}
Here, the harmonic vector field $\VHB_i$ becomes a Killing vector for $l=1$ and is exceptional.
\item[3.] For $K=0$, the exceptional mode exists only when $\K$ is isometric to $T^p\times \C^{n-p}$, where $\C^{n-p}$ is a Ricci flat space with no Killing vector. 
\Eitm

\subsubsection{Perturbation equations}

Vector perturbations of the metric and the energy-momentum tensor can be expanded in terms of the vector harmonics as
\Eqrsub{
&& h_{ab}=0,\quad
  h_{ai}=rf_a \VHB_i,\quad
  h_{ij}=2r^2 H_T \VHB_{ij},\\
&& \tau_{ab}=0,\ 
  \tau^a_{i}=r\tau_a \VHB_i,\ 
  \tau^i_{j}=\tau_T \VHB^i_{j}.
}
%
For the vector-type gauge transformation
\Eq{
\xi_a=0,\ \xi_i=rL \VHB_i
}
the perturbation variables transform as
\Eq{
\bar\delta f_a=-rD_a\Frac(L/r),\
\bar\delta H_T=\frac{k}{r}L,\
\bar\delta \tau_a=0,\
\bar\delta \tau_T=0.
}
Hence, we adopt the following combinations as the fundamental gauge-invariant variables for the vector perturbation:
\Eqr{
&\text{generic modes: } 
 & { \tau_a,\ \tau_T,\ F_a=f_a+\frac{r}{k}D_a H_T}
\\
&\text{exceptional modes:} 
  & { \tau_a,\ F^{(1)}_{ab}=rD_a\Frac(f_b/r)-rD_b\Frac(f_a/r)}
}
Note that for exceptional modes, $F_a=f_a$ because $H_T$ is not defined.

The reduced vector part of the Einstein equations come from the components corresponding to $G^a_i$ and $G^i_j$. In terms of the gauge-invariant variables defined above, these equations can be written as follows.
\Bitm
\item {\bf Generic modes}:
\Eqrsub{
&& \frac{1}{r^{n+1}}D^b\left(r^{n+1}F^{(1)}_{ab}\right)
      -\frac{k^2-(n-1)K}{r^2}F_a
    =-2{\bar\kappa}^2\tau_a, \notag\\
&&\label{kod:BulkPerturbationEq:vector1}
\\
&& \frac{k}{r^{n}}D_a(r^{n-1}F^a)=-{\bar\kappa}^2 \tau_T.
\label{kod:BulkPerturbationEq:vector2}
}
\item {\bf Exceptional modes}: $k^2=(n-1)K>0$. For these modes, the second of the above equations coming from $G^i_j$ does not exist.
\Eq{
\frac{1}{r^{n+1}}D^b\left(r^{n+1}F^{(1)}_{ab}\right)
  =-2{\bar\kappa}^2\tau_a.
\label{kod:BulkPerturbationEq:vector3}}
\Eitm

\subsection{Scalar Perturbation}

\subsubsection{Scalar harmonics}

Scalar functions on $\K$ can be expanded in terms of the harmonic functions defined by
\Eq{
(\hat \triangle + k^2)\SHB=0.
}
Correspondingly, scalar-type vector and tensor fields can be expanded in terms of harmonic vectors $\SHB_i$ and harmonic tensors $\SHB_{ij}$ define by
\Eqrsub{
&& \SHB_i=-\frac{1}{k}\hat D_i \SHB,\\
&& \SHB_{ij}=\frac{1}{k^2}\hat D_i\hat D_j \SHB +\frac{1}{n}\gamma_{ij}\SHB.
}
These harmonic tensors satisfy the following relations:
\Eqrsub{
&& \hat D_i \SHB^i=k\SHB,\\
&& [\hat \triangle + k^2-(n-1)K]\SHB_i=0,\\
&& \SHB^i_i=0,\quad
\hat D_j \SHB^j_i=\frac{n-1}{n}\frac{k^2-nK}{k}\SHB_i,
\\
&& [\hat \triangle +k^2-2nK]\SHB_{ij}=0.
}

Note that as in the case of vector harmonics, there are some exceptional modes:
\Bitm
\item[i)] $k=0$: $\SHB_i\equiv0$, $\SHB_{ij}\equiv0$.
\item[ii)] $k^2=nK$ ($K>0$): $\SHB_{ij}\equiv0$.
\Eitm
%

For scalar harmonics, $k^2=0$ is obviously always the allowed lowest eigenvalue. Therefore, the information on the second eigenvalue is important. In general, it is difficult to find such information. However,  when $\K^n$ is a compact Einstein space with $K>0$, we can obtain a useful constraint as follows. Let us define $Q_{ij}$ by
\Eqn{
Q_{ij}:=D_i D_j Y -\frac{1}{n}g_{ij}\triangle Y.
}
Then, we have the identity
\Eqn{
Q_{ij}Q^{ij}=D^i(D^i Y D_iD_j Y-YD_i\triangle Y-R_{ij}D^jY)
  +Y\insbra{\triangle(\triangle +(n-1)K)}Y
  -\frac{1}{n}(\triangle Y)^2.
}
For $Y=\SHB$, integrating this identity, we obtain the constraint on the second eigenvalue
\Eq{
k^2\ge nK.
}
For $\K^n=S^n$, the equality holds because the full spectrum is given by
\Eq{
k^2=\ell(\ell+n-1),\quad(\ell=0,1,2,\cdots).
}
%

\subsubsection{Perturbation equations}

The scalar perturbation of the metric and the energy-momentum tensor can be expanded as
\Eqrsub{
&& h_{ab}=f_{ab}\SHB,\ 
   h_{ai}=rf_a \SHB_i,\ 
   h_{ij}=2r^2(H_L\gamma_{ij}\SHB+H_T \SHB_{ij}),\\
&& \tau_{ab}=\tau_{ab}\SHB,\ 
   \tau^a_{i}=r\tau_a \SHB_i,\ 
   \tau^i_{j}=\delta {\bar P} \delta^i_j \SHB + \tau_T \SHB^i_j.
}
%
For the scalar-type gauge transformation
\Eq{
\xi_a=T_a \SHB,\ \xi_i=rL \SHB_i,
}
these harmonic expansion coefficients for generic modes $k^2(k^2-nK)>0$ of a scalar-type perturbation  transform as
\Eqrsub{
&& \bar\delta f_{ab}=-D_a T_b -D_b T_a,\\
&& \bar\delta f_a=-rD_a\Frac(L/r)+\frac{k}{r}T_a,\\
&& \bar\delta H_L=-\frac{k}{nr}L-\frac{D^ar}{r}T_a,\\
&& \bar\delta H_T=\frac{k}{r}L,\\
&& \bar\delta \tau_{ab}=-T^cD_c{T}_{ab}
   -{\bar T}_{ac}D_bT^c-{T}_{bc}D_aT^c,\\
&& \bar\delta \tau_a=\frac{k}{r}({T}_{ab}T^b-{P}T_a),\\
&& \bar\delta (\delta {P})=-T^aD_a {P},\\ 
&& \bar\delta \tau_T=0.
}

From these we obtain
\Eq{
\bar\delta X_a=T_a;\quad
X_a=\frac{r}{k}\left(f_a+\frac{r}{k}D_a H_T\right).
}
Hence, the fundamental gauge invariants can be given by $\tau_T$ and the following combinations:
\Eqrsub{
&& F=H_L+\frac{1}{n}H_T+\frac{1}{r}D^ar X_a,\\
&& F_{ab}=f_{ab}+D_aX_b+D_bX_a,\\
&& \Sigma_{ab}=\tau_{ab}+{\bar T}^c_b D_aX_c
              +{\bar T}^c_aD_bX_c+X^cD_c{\bar T}_{ab},\\
&& \Sigma_a=\tau_a -\frac{k}{r}({\bar T}^b_a X_b - {\bar P}X_a),\\
&& \Sigma_L = \delta {\bar P} +X^aD_a{\bar P}.
}
%

The scalar part of the Einstein equations comes from $G_{ab}$, $G_{ai}$ and $G^i_j$. First, from $\delta G_{ab}$, we obtain
\Eqr{
&&-\square F_{ab}+D_aD_c F^c_b+D_bD_cF^c_a 
  +n\frac{D^cr}{r}(-D_cF_{ab}+D_aF_{cb}+D_bF_{ca}) 
\notag\\
&&\quad +\,{}^m\!R^c_aF_{cb}  +\,{}^m\!R^c_bF_{ca}-2\,{}^m\!R_{acbd}F^{cd}
   +\left(\frac{k^2}{r^2}-{\bar R}+2{\bar\Lambda}\right)F_{ab}
   -D_aD_b F^c_c
\notag\\
&&\quad -2n\left(D_aD_bF+\frac{1}{r}D_arD_bF
         +\frac{1}{r}D_brD_aF \right) 
\notag\\
&& -\left[ D_cD_dF^{cd}+\frac{2n}{r}D^cr D^dF_{cd}\right.
   +\Big(\frac{2n}{r}D^cD^dr
        +\frac{n(n-1)}{r^2}D^crD^dr 
\notag\\
&&\quad 
  -{}^m\!R^{cd}\Big)F_{cd}
  -2n\square F -\frac{2n(n+1)}{r}Dr\cdot DF
           +2(n-1)\frac{k^2-nK}{r^2}F 
\notag\\
&&\quad  \left. -\square F^c_c-\frac{n}{r}Dr\cdot DF^c_c
          +\frac{k^2}{r^2}F^c_c \right]g_{ab} 
          =2{\bar\kappa}^2 \Sigma_{ab}.
\label{kod:BulkPerturbationEq:scalar1}
}
Second, from $\delta G^a_i$, we obtain
\Eq{
\frac{k}{r}\left[-\frac{1}{r^{n-2}}D_b(r^{n-2}F^b_a)
    +rD_a\left(\frac{1}{r}F^b_b\right)+2(n-1)D_aF\right] 
 =2{\bar\kappa}^2\Sigma_a.
\label{kod:BulkPerturbationEq:scalar2}
}
Finally, from the trace-free part of $\delta G^i_j$, we obtain
\Eq{
-\frac{k^2}{2r^2}\left[2(n-2)F+ F^a_a\right]={\bar\kappa}^2 \tau_T,
\label{kod:BulkPerturbationEq:scalar3}
}
and from the trace $\delta G^i_i$,
\Eqr{
&&-\frac{1}{2}D_aD_b F^{ab}-\frac{n-1}{r}D^arD^bF_{ab} 
\notag\\
&& +\left[\frac{1}{2}{}^m\!R^{ab}
     -\frac{(n-1)(n-2)}{2r^2}D^arD^br-(n-1)
            \frac{D^aD^br}{r}\right]F_{ab} 
\notag\\
&&\quad  +\frac{1}{2}\square F^c_c+\frac{n-1}{2r}Dr\cdot DF^c_c
         -\frac{n-1}{2n}\frac{k^2}{r^2}F^c_c 
   +(n-1)\square F 
\notag\\
&&\quad 
   + \frac{n(n-1)}{r}Dr\cdot DF
   -\frac{(n-1)(n-2)}{n}\frac{k^2-nK}{r^2}F 
    ={\bar\kappa}^2\Sigma_L.
\label{kod:BulkPerturbationEq:scalar4}
}

Note that for the exceptional mode with $k^2=nK>0$, the third equation does not exist, and for the mode with $k^2=0$, the second and the third equations do not exist. For these exceptional modes, the other equations hold without change, but the variables introduced in the above are not gauge invariant.

Although the energy-momentum conservation equation $\nabla_N T^N_M=0$ can be derived from the Einstein equations, it is often useful to know its explicit form. For scalar-type perturbations, they are given by the following two sets of equations:
\Eqrsub{
&& \frac{1}{r^{n+1}}D_a(r^{n+1}\Sigma^a)
  -\frac{k}{r}\Sigma_L +\frac{n-1}{n}\frac{k^2-nK}{kr}\tau_T 
\nonumber\\
&& \qquad +\frac{k}{2r}({\bar T}^{ab}F_{ab}-{\bar P}F^a_a)=0,
\label{kod:BulkPerturbationEq:EM1}\\
&& \frac{1}{r^n}D_b\left[r^n(\Sigma^b_a-{\bar T}^c_a F^b_c)\right]
   +\frac{k}{r}\Sigma_a-n\frac{D_a r}{r}\Sigma_L
\nonumber\\
&& \quad +{\bar T}_a^b D_bF-{\bar P}D_aF
   +\frac{1}{2}\left({\bar T}^b_a D_bF^c_c
   -{\bar T}^{bc}D_a F_{bc}\right)=0.
\label{kod:BulkPerturbationEq:EM2}
}
%

\section{Stability of Static Black Holes}
\label{kod:sec:SBH}

We study the stability of static black holes utilising the gauge-invariant formulation for perturbations explained in the previous section. We consider the static Einstein black hole which corresponds to the case with $m=2$ of the general background considered in the previous section and has the metric \eqref{kod:background:sbh}. The key point is the fact that gauge-invariant perturbation equations can be reduced to decoupled single master equations of the Schr\"odinger type for any type of perturbations in this background.

\subsection{Tensor Perturbations}

The gauge-invariant equation for tensor perturbations is already given by a single equation for each mode, Assuming that the source term vanishes, it reads
\Eq{
-\pd_t H_T^2 + f\pd_r(f\pd_r H_T) - \frac{k^2+2K}{r^2}f H_T=0.
}
Here, note that even if there exist electromagnetic fields, $\tau_T$ vanishes because the electromagnetic field is vector-like and does not produce a tensor-type quantity in the linear order at least.

With the help of the Fourier transformation with respect to $t$, i.e., assuming $H_T \propto e^{-i\omega t}$, this equation can be put into the Schr\"odinger-type eigenvalue problem;
\Eqr{
&& H_T=  r^{-n/2} \Phi(r) e^{-i\omega t},\\
&& \omega^2\Phi = -f\pd_r(f\pd_r \Phi) + V_t \Phi
}
where
\Eqr{
V_t &=& \frac{f}{r^2}\insbra{ k^2+2K + \frac{nrf'}{2}+\frac{n(n-2)f}{4}}
\\
  &=& \frac{f}{r^2}\insbra{k^2+\frac{n^2-2n+8}{4}K
   -\frac{n(n+2)}{4}\lambda r^2 + \frac{n^2M}{2r^{n-1}}
    -\frac{n(3n-2)Q^2}{4r^{2n-2}} }. \notag
}
If $V_t$ is non-negative, we can directly conclude the stability. However, it is not so easy to see whether $V_t$ is non-negative or not outside the horizon. This technical difficulty is easily resolved by considering the energy integral
\Eq{
E:= \int_{r_h}^{r_\infty} dr \insbra{\frac{1}{f}(\pd_t H_T)^2 + f (\pd_r H_T)^2 
  +\frac{k^2+ 2K}{r^2} H_T^2 }.
}
From the equation for $H_T$, we find that
\Eq{
\pd_t E= 2 \insbra{f \pd_t H_T \pd_r H_T}_{r_h}^{r_\infty}=0.
}
Hence, in the case $\K$ is a constant curvature space, the condition on the spectrum $k^2\ge n|K|$ guarantees the positivity of all terms in $E$, and as a consequence the stability of the system.

\subsection{Vector Perturbations}

\subsubsection{Master equation}

For vector perturbations, the energy-momentum conservation law is written
\Eq{
D_a(r^{n+1}\tau^a)+\frac{m_v}{2k}r^n\tau_T=0.
\label{kod:EMconservation:vector}
}
For $m_v\equiv k^2-(n-1)K\not=0$, with the help of this equation, the second of the perturbation equations, \eqref{kod:BulkPerturbationEq:vector2}, can be written
\Eq{
D_a(r^{n-1}F^a)=\frac{2\kappa^2}{m_v} D_a(r^{n+1}\tau^a).
}
In the case of $m=2$, from this it follows that $F^a$ can be written in terms of a 
variable $\Omega$ as
\Eq{
r^{n-1}F^a=\epsilon^{ab}D_b \Omega
    +\frac{2\kappa^2}{m_v}r^{n+1}\tau^a.
\label{kod:FaByOmega}
}
Further, the first of the perturbation equations, \eqref{kod:BulkPerturbationEq:vector1}, is equivalent to 
\Eq{
D_a\left( r^{n+1}F^{(1)} \right)
  -m_v r^{n-1}\epsilon_{ab}F^b
  =-2\kappa^2 r^{n+1}\epsilon_{ab}\tau^b,
\label{kod:BasicEq:metric:vector1}
}
where $\epsilon_{ab}$ is the two-dimensional Levi-Civita tensor for 
$g_{ab}$, and
\Eq{
F^{(1)}=\epsilon^{ab}r D_a\pfrac{F_b}{r}
       =\epsilon^{ab}r D_a\pfrac{f_b}{r}.
\label{kod:F^(1):def}
}
Inserting the expression for $F_a$ in terms of $\Omega$ into \eqref{kod:BasicEq:metric:vector1}, we obtain the master equation
\Eq{
r^nD_a\left( \frac{1}{r^n}D^a \Omega \right)
  -\frac{m_v}{r^2} \Omega 
  =-\frac{2\kappa^2}{m_v} r^n\epsilon^{ab}D_a(r\tau_b).
\label{kod:MasterEq:vector:metric}
}

Next, for $m_v=0$, the perturbation variables $H_T$ and 
$\tau_T$ do not exist. The matter variable $\tau_a$ is still 
gauge-invariant, but concerning the metric variables, only the 
combination $F^{(1)}$ defined in terms of $f_a$ in \eqref{kod:F^(1):def} 
is gauge invariant. In this case, the Einstein equations are reduced 
to the single equation \eqref{kod:BasicEq:metric:vector1}, and the 
energy-momentum conservation law is given by 
\eqref{kod:EMconservation:vector} without the $\tau_T$ term. 
Hence, 
$\tau_a$ can be expressed in terms of a function $\tau^{(1)}$ as
\Eq{
r^{n+1} \tau_a=\epsilon_{ab}D^b\tau^{(1)}.
\label{kod:tau1:def}
}
Inserting this expression into \eqref{kod:BasicEq:metric:vector1} with 
$\epsilon^{cd}D_c(F_d/r)$ replaced by $F^{(1)}/r$, we obtain
\Eq{
D_a(r^{n+1}F^{(1)})=-2\kappa^2D_a\tau^{(1)}.
} 
Taking account of the freedom of adding a constant in the definition 
of $\tau^{(1)}$, the general solution can be written
\Eq{
F^{(1)}=-\frac{2\kappa^2 \tau^{(1)}}{r^{n+1}}.
\label{kod:BasicEq:vector:metric:exceptional}
}
Hence, there exists no dynamical freedom in these special modes. In 
particular, in the source-free case in which $\tau^{(1)}$ is a 
constant and $K=1$, this solution corresponds to adding a small 
rotation to the background static black hole solution.

\subsubsection{Neutral black holes}

For a neutral static Einstein black hole, the master equation for a generic mode can be put into the canonical form as
\Eqr{
&& \Omega = r^{n/2}\Phi(r) e^{-i\omega t},\\
&& \omega^2\Phi = -f\pd_r(f\pd_r \Phi) + V_v \Phi
}
where
\Eqr{
V_v &=& \frac{f}{r^2}\insbra{ m_v - \frac{nrf'}{2}+\frac{n(n+2)f}{4}} \notag\\
    &=& \frac{f}{r^2}\insbra{ k^2 + \frac{n(n+2)K}{4}-\frac{n(n-2)}{4}\lambda r^2
    -\frac{3n^2 M}{2r^{n-1}} }.
}

This equation is identical to the Regge-Wheeler equation for $n=2, K=1$ and $\lambda=0$. In this case, we can put $V_v$ into an obviously non-negative form as
\Eq{
V_v =\frac{f}{r^2}\inpare{m_v + 3f},
}
proving the stability of the black hole against vector perturbations (or axial or odd perturbations).

\begin{figure}
\begin{minipage}[b]{5cm}   
   \centerline{\includegraphics[width=5 cm]{\FigDir/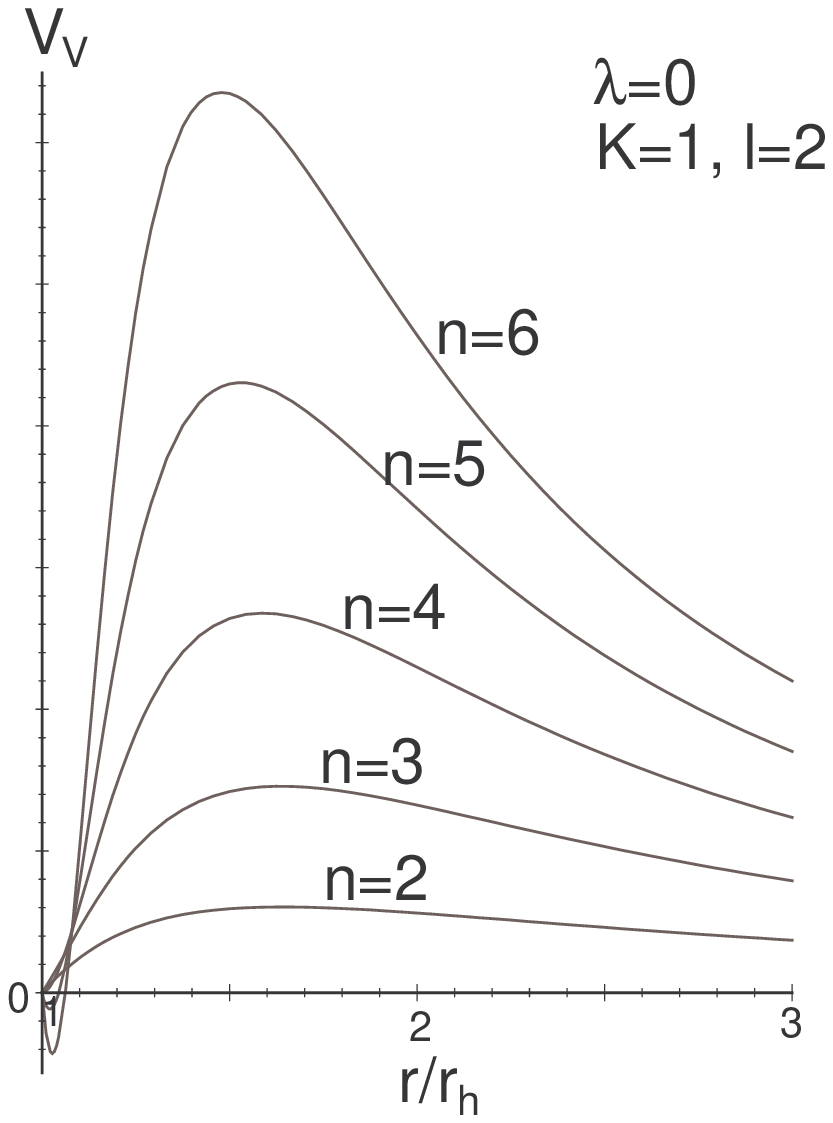}}
   \caption{$V_v$ for $K=1,\lambda=0,l=2$.}
   \label{kod:fig:PotVL0l2}
\end{minipage}
\hspace{5mm}
\begin{minipage}[b]{5cm}   
   \centerline{\includegraphics[width=5 cm]{\FigDir/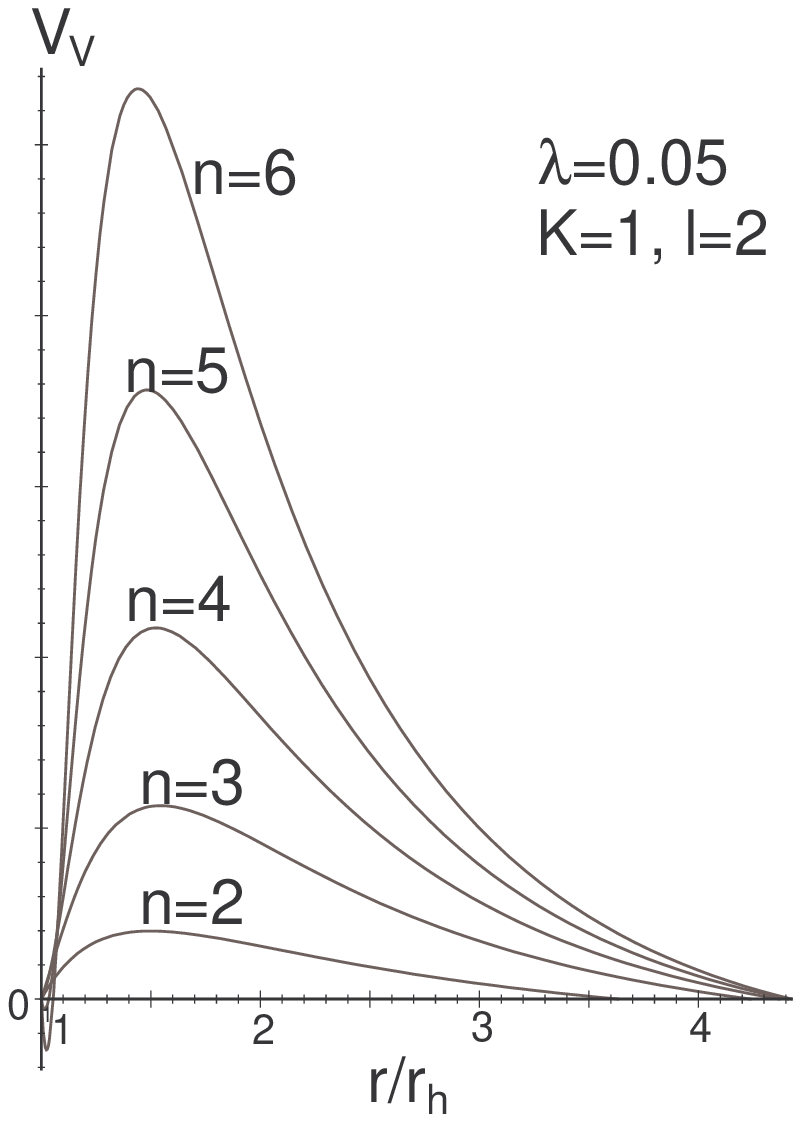}}
   \caption{$V_v$ for $K=1,\lambda>0, l=2$.}
   \label{kod:fig:PotVPLl2}
\end{minipage}
\hspace{5mm}
\begin{minipage}[b]{5cm}   
   \centerline{\includegraphics[width=5 cm]{\FigDir/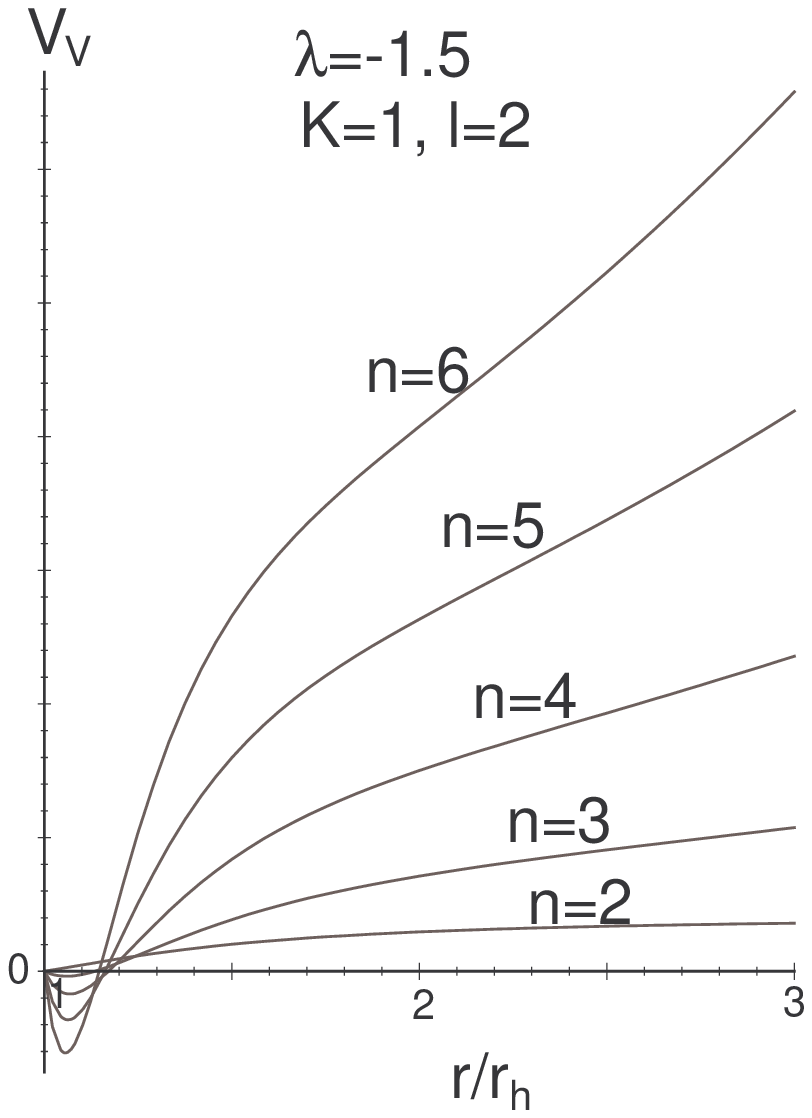}}
   \caption{$V_v$ for $K=1,\lambda<0, l=2$.}
   \label{kod:fig:PotVNLl2}
\end{minipage}
\end{figure}

In higher dimensions, the potential $V_v$ is not positive definite anymore and we can not use this type of argument. However, we can still prove the stability with the help of the conserved energy integral as in the case of tensor perturbations. In the present case, if we define $E$ as
\Eq{
E:= \int_{r_h}^{r_\infty} \frac{dr}{r^n} \insbra{ \frac{1}{r}(\pd_t\Omega)^2
   + f(\pd_r\Omega)^2 + \frac{m_v}{r^2}\Omega^2 },
}
we have
\Eq{
\dot E =2\insbra{ \frac{f}{r^n} \pd_t \Omega \pd_r \Omega}_{r_h}^{r_\infty}=0.
}
Further, all terms of $E$ is non-negative because $m_v\ge0$. Hence, the stability can be concluded.


\subsubsection{Charged black hole}

The formulation for neutral static black holes can be extended to charged static black holes. The final master equations consist of two equations: the extension of the equation for gravitational perturbations with an electromagnetic source and the equation coming from the Maxwell equations\cite{Kodama.H&Ishibashi2004}:
\Eqrsub{
&& r^nD_a\left( \frac{1}{r^n}D^a\Omega \right)
  -\frac{m_v}{r^2}\Omega 
    =\frac{2\kappa^2 q}{r^2}\A,\\
&&\frac{1}{r^{n-2}}D_a(r^{n-2}D^a\A)-\frac{m_v+2(n-1)K}{r^2}\A 
  =\frac{q}{r^{2n}}\left[ m_v \Omega+2\kappa^2 q \A \right],
}
where $\A$ is the gauge invariant representing a vector perturbation of the vector potential of the electromagnetic field defined by
\Eq{
\delta A_a=0, \delta A_i=\A \VHB_i,
}
and  $q$ is the black hole charge related to the charge parameter $Q$ in the background metric by
\Eq{
Q^2:=\frac{\kappa^2 q^2}{n(n-1)},
}

By taking appropriate combinations, these equations can be transformed to the two decoupled equations 
\Eq{
-\partial_t^2 \Phi_\pm=(-\partial_{r_*}^2+V_\pm)\Phi_\pm,
}
where the effective potentials are given by
\Eq{
V_\pm=\frac{f}{r^2}\left[m_v +\frac{n(n+2)K}{4}
   -\frac{n(n-2)}{4}\lambda r^2+\frac{n(5n-2)Q^2}{4r^{2n-2}} 
   +\frac{\mu_\pm}{r^{n-1}}\right],
}
with
\Eqr{
&& \mu_\pm=-\frac{n^2+2}{2}M \pm \Delta;\\
&& \Delta^2= (n^2-1)^2M^2+2n(n-1)m_v Q^2.
}
%

\subsubsection{S-deformation}

The effective potentials $V_\pm$ are not positive definite as in the neutral case. In the present case, we prove that the system is still stable not by the energy integral method, but rather by a different method, which we call {\em the S-deformation}\cite{Ishibashi.A&Kodama2003}. 

We first explain the basic idea by the eigen value equation
\Eq{
\omega^2\Phi=\inpare{-D^2+ V(r)}\Phi,
}
where $D=\pd_{r_*}$. If there exists an unstable mode with $\omega^2<0$, we can show $\Phi$ falls off sufficiently rapidly at horizon and at infinity if $V$ is non-negative at horizon and at infinity. Hence, we obtain the integral identity,
\Eq{
\omega^2 \int_{r_h}^{r_\infty} |\Phi|^2 \frac{dr}{f}
= \int_{r_h}^{r_\infty} \insbra{|D\Phi|^2+ V(r)|\Phi|^2} \frac{dr}{f}.
}
If $V(r)$ is non-negative definite, this leads to contradiction and hence proves the stability because the right-hand side is non-negative. In contrast, in the case in which the sign of $V$ is not definite, we cannot say anything about stability from this equation. 

In order to treat such a case, let us replace $D$ by $D=\t D -S$. Then, by partial integrations, we obtain the modified integral identity with $D$ and $V$ replaced by $\t D$ and $\t V$ given by
\Eq{
\t V= V + f\frac{d S}{dr}-S^2.
}
Hence, if we can find $S$ such that the modified effective potential $\t V$ is non-negative, we can establish the stability of the system even when the original potential is not non-negative definite.

For example, by the S-transformation with
\Eq{
S=\frac{nf}{2r},
}
the effective potentials $V_\pm$ above can be modified into
\Eqr{
&\tilde V_\pm &= V_\pm +f\frac{dS}{dr}-S^2 \notag\\
&&=\frac{f}{r^2}\left[ m_v
   +\frac{1}{r^{n-1}}\left( \frac{3n^2}{2}M +\mu_\pm \right) \right].
}
Here, $\t V_+$ is obviously positive definite. We can also show that $\t V_-$ is also positive definite. Hence, a charged static Einstein black hole is stable for vector perturbations.

\subsection{Scalar Perturbations}

\subsubsection{Master equation}

For a static Einstein black hole background, assuming that $F_{ab}, F\propto e^{-i\omega t}$,  we can reduce the whole linearised Einstein equations into a single master equation, as in the case of vector perturbations\cite{Kodama.H&Ishibashi2003}:
\Eq{
\omega^2\Phi = -f\pd_r(f\pd_r \Phi) + V_s\Phi,
}
where the master variable $\Phi$ is defined as
\Eqr{
&& \Phi=\frac{nr^{n/2}}{H}\left( 2F+\frac{F^r_t}{i\omega r} 
\right);
\label{kod:BH:ScalarP:MasterVar}\\
&& H=m+\frac{n(n+1)M}{r^{n-1}},\quad  m=k^2-nK,
}
and the effective potential $V_s$ is given by
\Eqr{
V_s(r) &=& \frac{f U(r)}{16r^2H^2};\\
U(r) &=&
 -\left[n^3(n+2)(n+1)^2x^2-12n^2(n+1)(n-2)mx\right.
 \notag\\
&& \left. \qquad +4(n-2)(n-4)m^2\right] \lambda r^2  +n^4(n+1)^2x^3 \notag\\
&& +n(n+1)\left[4(2n^2-3n+4)m+n(n-2)(n-4)(n+1)K\right]x^2
   \notag\\
&& -12n\left[(n-4)m+n(n+1)(n-2)K \right]mx \notag\\
&&   +16m^3+4Kn(n+2)m^2,
\label{kod:Q:NonStatic}
}
with $x=2M/r^{n-1}$.

\subsubsection{Neutral black holes}

The above master equation is identical to the Zerilli equation for the four-dimensional Schwarzschild black hole ($n=2, K=1$ and $\lambda=0$). In this case, from
\Eq{
V_s=\frac{f}{r^2H^2}\left( m^2(m+2)+ \frac{6m^2M}{r}
      +\frac{36mM^2}{r^2}+\frac{72M^3}{r^3} \right) \ge0,
}
where $m=(l-1)(l+2)$($l=2,3,\cdots$), we can easily prove the stability of the black hole. In higher dimensions, however, the effective potential $V_s$ is not positive definite. Hence, an instability may arise.

\begin{figure}
\begin{minipage}[b]{5cm}   
   \centerline{\includegraphics[width=5 cm]{\FigDir/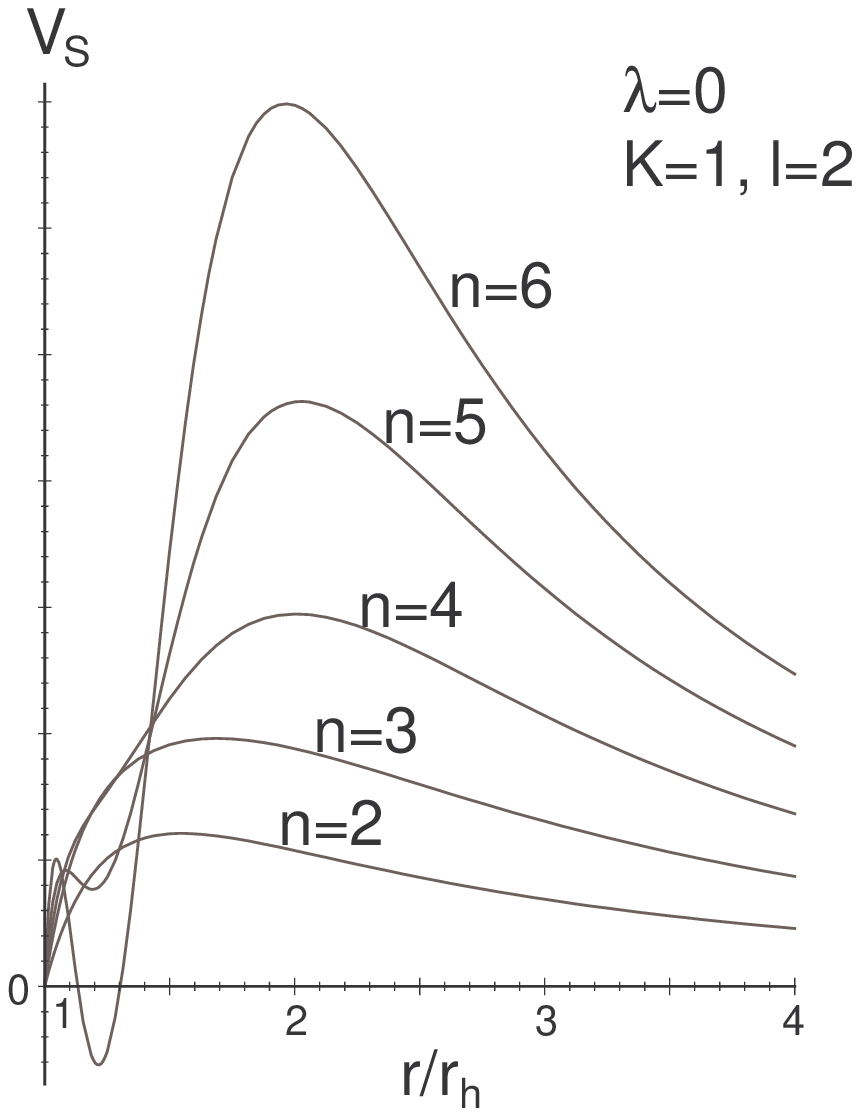}}
   \caption{$V_s$ for $K=1,\lambda=0,l=2$.}
   \label{kod:fig:PotSL0l2}
\end{minipage}
\hspace{5mm}
\begin{minipage}[b]{5cm}   
   \centerline{\includegraphics[width=5 cm]{\FigDir/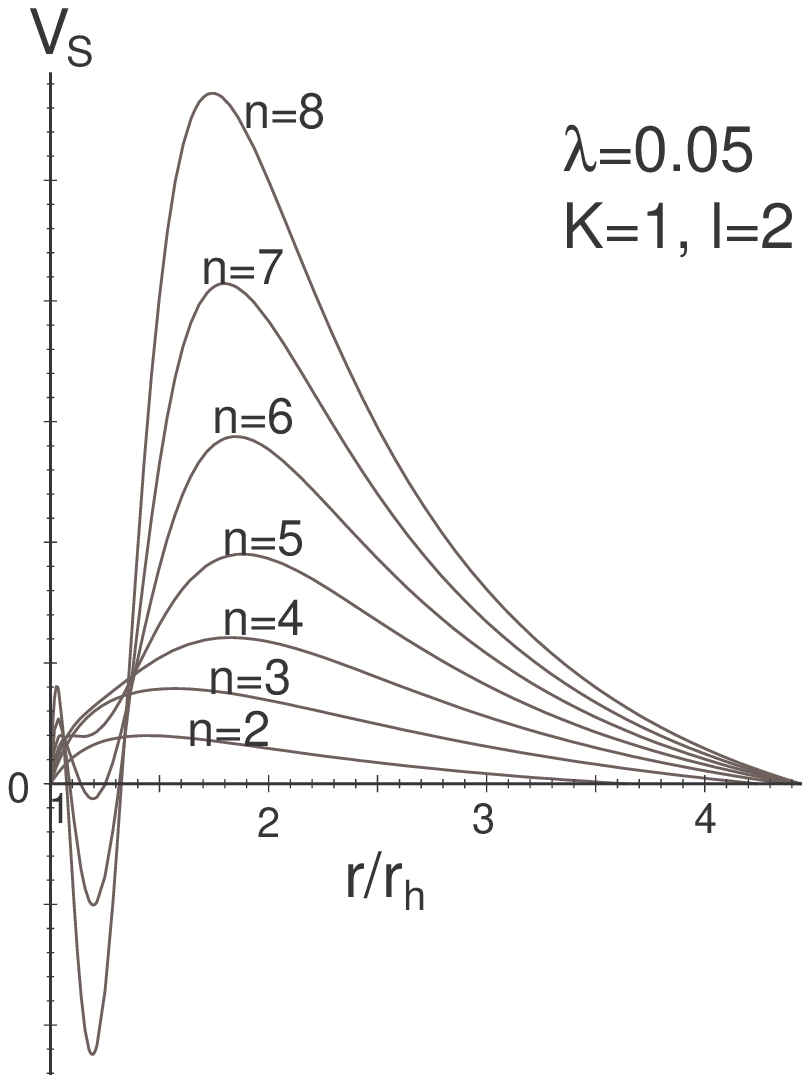}}
   \caption{$V_s$ for $K=1,\lambda>0, l=2$.}
   \label{kod:fig:PotSPLl2}
\end{minipage}
\hspace{5mm}
\begin{minipage}[b]{5cm}   
   \centerline{\includegraphics[width=5 cm]{\FigDir/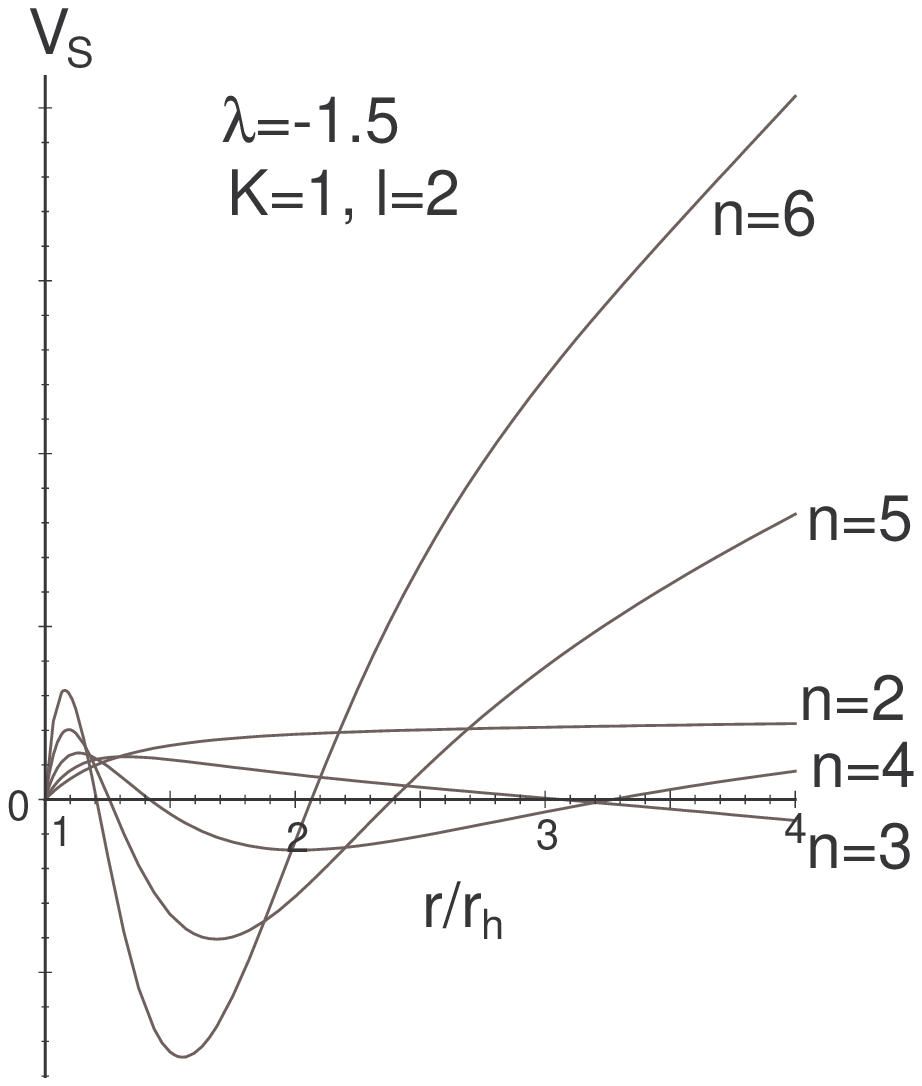}}
   \caption{$V_s$ for $K=1,\lambda<0, l=2$.}
   \label{kod:fig:PotSNLl2}
\end{minipage}
\end{figure}

Nevertheless, in the case of $K=1$ and $\lambda=0$, i.e. for the Schwarzschild-Tangherlini black hole, we can prove the stability by applying the S-deformation to the energy integral. First, from the above master equation, we obtain
\Eqr{
&& E:= \int_{r_0}^{r_\infty} \frac{dr}{f}\insbra{(\pd_t\Phi)^2 + (D\Phi)^2
   + V_s \Phi^2 } ,\\
&& \dot E = \insbra{2f \pd_t \Phi\pd_r\Phi}_{r_h}^{r_\infty} =0,
}
where $D=f\pd_r$. Next, we replace $D$ to
\Eq{
\t D=f\pd_r + S.
}
Then, by partial integration we obtain
\Eq{
E=\int_{r_0}^{r_\infty} \frac{dr}{f}\insbra{(\pd_t\Phi)^2 + (\t D\Phi)^2
   + \t V_s \Phi^2 },
}
where 
\Eq{
\t V_s = V_s + f\frac{dS}{dr}-S^2.
}

For example, for
\Eq{
  S = \frac{f}{h}\frac{dh}{dr} \,, \quad 
  h \equiv r^{n/2+l-1} \left\{ 
                              (l-1)(l+n) + n(n+1)x /2  
                       \right\} \,.   
}
we obtain
\Eq{
\tilde V_s = \frac{f(r) \tilde Q(r)}{4r^{2}\inrbra{(l-1)(l+n) + n(n+1) x/2 }}\,, 
}  
where   
\Eq{
 \tilde Q(r) \equiv lx[ ln (n+1)x +2(l-1)\{n^2+n(3l-2) + (l-1)^2\}] \,.
}
Clearly $\tilde V_s > 0$.

\subsubsection{Charged black holes}

For charged black holes, we can also reduce the perturbation equations to decoupled single master equations. First, we generalise the master variable $\Phi$ for the metric perturbation given in \eqref{kod:BH:ScalarP:MasterVar} by replacing $H$ by 
\Eq{
H=m+\frac{n(n+1)M}{r^{n-1}}-\frac{n^2 Q^2}{r^{2n-2}}.
}
Next, we introduce the gauge-invariant variable $\A$ in terms of which the scalar perturbation of the electromagnetic field is expressed as
\Eqrsub{
&& \delta \F_{ab}+D_c(E_0X^c)\epsilon_{ab}\SHB=\E \epsilon_{ab}\SHB,\\
&& \delta\F_{ai}-kE_0\epsilon_{ab}X^b\SHB_i=r\epsilon_{ab}\E^b\SHB_i,\\
&& \delta\F_{ij}=0,
}
with
\Eq{
\E_a =\frac{k}{r^{n-1}} D_a \A,\quad
r^n\E = -k^2\A + \frac{q}{2}(F^c_c-2nF).
}

Then, the Einstein and Maxwell equations for scalar perturbations of a charge Einstein black hole can be reduced to the following two coupled equations\cite{Kodama.H&Ishibashi2004}:
\Eqrsub{
& \omega^2\Phi=
  & -\frac{d^2\Phi}{dr_*^2}+V_s\Phi
  +\frac{\kappa^2q fP_{S1}}{r^{3n/2}H^2}\A,\\
& \omega^2\A=
  & -r^{n-2}\frac{d}{dr_*}
    \left(\frac{1}{r^{n-2}}\frac{d\A}{dr_*} \right)
    +f\left(\frac{k^2}{r^2}\A+\frac{2n^2(n-1)^2Q^2f}{r^{2n}H}\right)\A
    \notag\\
&&\quad +f\frac{(n-1)q}{r^{n/2}}\left( 
       \frac{4H^2-nP_Z}{4nH}\Phi+fr\partial_r\Phi \right).
}
where  $V_s$ is the extension to the charged case of the corresponding potential in the neutral case, 
\Eqr{
&& V_s=\frac{f(r)U_s(r)}{16r^2H^2};\\
&& U_s=\left[-n^3 (n+2) (n+1)^2 x^2+4n^2(n+1)\{n(n^2+6 n-4) z
      +3(n-2) m\} x \right.\notag\\
&&\quad\left. -12 n^5 (3 n-2) z^2-8 n^2 (11 n^2-26 n+12) m z
      -4 (n-2) (n-4) m^2\right] \lambda r^2 \notag\\
&&\quad  +n^4 (n+1)^2 x^3+n(n+1)\left\{-3 n^2 (5 n^2-5 n+2) z  
    +4 (2 n^2-3 n+4) m  \right.\notag\\
&&\quad \left. +n (n-2) (n-4) (n+1) K\right\} x^2
    +4n\left[n^2 (4 n^3+5 n^2-10 n+4) z^2 \right.\notag\\
&&\quad -\left\{n (34-43 n+21 n^2) m
        + n^2 (n+1) (n^2-10 n+12) K\right\} z
  \notag\\
&&\quad \left.-3 (n-4) m^2-3 n (n+1) (n-2) K m\right] x
        -4 n^5 (3 n-2) z^3
\notag\\
&&\quad +12n^2\left\{2(-6 n+3 n^2+4) m+n^2 (3 n-4) (n-2) K\right\} z^2
\notag\\ 
&&\quad +\left\{4 (13 n-4) (n-2) m^2
                 +8 n^2 (11 n^2-18 n+4) K m\right\} z
\notag\\
&&\quad +16 m^3+4 n (n+2) K m^2,
}
with $z=Q^2/r^{2(n-1)}$, and  $P_{S1}$ and $P_Z$ are the functions of $r$ given by
\Eqr{
& P_{S1}=& \left[ -4n^4z+2n^2(n+1)x-4n(n-2)m \right]\lambda r^2 \notag\\
&& +\left\{ 2n^2(n-1)x+4n(n-2)m+4n^3(n-2)K \right\}z
 \notag\\
&& -n^2(n^2-1)x^2+\left\{ -4n(n-2)m+2n^2(n+1)K \right\}x
\notag\\
&& +4m^2+4n^2mK, \\
&P_Z=& \left[ -n^2(n+1)x+2n^2(3n-2)z+2(n-2)m \right]\lambda r^2 \notag\\
&& +n(n+1)x^2+\left\{ n^2(3n-7)z+(4n-2)m+n(n+1)(n-2)K \right\}x
\notag\\
&& -2n^2(n-2)z^2-\left\{ (6n-4)m+2n^2(3n-4)K \right\}z 
    -2nmK.
}

As in the case of vector perturbations, we can find linear combinations of $\A$ and $\Phi$, in terms of which these equations are transformed to the decoupled equations
\Eq{
\frac{\omega^2}{f}\Phi_\pm=-(f\Phi_\pm')'+\frac{V_\pm}{f}\Phi_\pm;\quad
 V_\pm =\frac{fU_\pm}{64r^2H_\pm^2},
}
Here,
\Eq{
H_+=1-\frac{n(n+1)}{2}\delta x,\quad
H_-=m+\frac{n(n+1)}{2}(1+m\delta)x,\\
}
and $\delta$ is a non-negative constant determined from $Q$ by
\Eq{
Q^2=(n+1)^2M^2 \delta( 1+m\delta).
}

The effective potentials $U_\pm$ can be expressed in terms of $x,\lambda r^2, m$ and $\delta$ as follows:
\Eqr{
& U_+ =
& \left[-4 n^3 (n+2) (n+1)^2 \delta^2 x^2-48 n^2 (n+1) (n-2) \delta x
\right.\notag\\
&&\left.   -16 (n-2) (n-4)\right] \lambda r^2
  -\delta^3 n^3 (3 n-2) (n+1)^4 (1+m \delta) x^4
\notag\\
&&   +4 \delta^2 n^2 (n+1)^2 
   \left\{(n+1)(3n-2) m \delta+4 n^2+n-2\right\} x^3
\notag\\   
&&   +4 \delta (n+1)\left\{
   (n-2) (n-4) (n+1) (m+n^2 K) \delta-7 n^3+7 n^2-14 n+8
   \right\}x^2
\notag\\   
&&  + \left\{16 (n+1) \left(-4 m+3 n^2(n-2) K\right) \delta
     -16 (3 n-2) (n-2) \right\}x
\notag\\   
&&    +64 m+16 n(n+2) K,
}
\Eqr{
& U_- =
  & \left[-4 n^3 (n+2) (n+1)^2 (1+m \delta)^2 x^2
      +48 n^2 (n+1) (n-2) m (1+m \delta) x  \right.
\notag\\
&& \left.  -16 (n-2) (n-4) m^2\right] y
     -n^3 (3 n-2) (n+1)^4 \delta (1+m \delta)^3 x^4
\notag\\
&& -4 n^2 (n+1)^2 (1+m \delta)^2 
     \left\{(n+1)(3 n-2) m \delta-n^2\right\} x^3
\notag\\  
&&  +4 (n+1) (1+m \delta)\left\{ m (n-2) (n-4) (n+1) (m+n^2 K) \delta
  \right. \notag\\
&& \left. \quad  +4 n (2 n^2-3 n+4) m+n^2 (n-2) (n-4) (n+1)K \right\}x^2
\notag\\
&&  -16m \left\{ (n+1) m \left(-4 m+3 n^2(n-2) K\right) \delta
\right.\notag\\
&&\left.  +3 n (n-4) m+3 n^2 (n+1) (n-2)K \right\}x
\notag\\
&&      +64 m^3+16 n(n+2)m^2 K.
}

\begin{table}[t]
\caption{{\bf stability of generalised static black holes.} }
\label{kod:tbl:stability}
\begin{center}\large
\begin{tabular}{|l|l|c|c|c|c|}
\hline\hline
\multicolumn{2}{|c|}{}& {Tensor}
  & {Vector}& \multicolumn{2}{c|}{Scalar}\\
\cline{5-6}
\multicolumn{2}{|c|}{}&$\forall Q$ & $\forall Q$ 
&$Q=0$ & $Q\not=0$ \\
\hline
$K=1$& $\lambda=0$ & OK & OK  
     & OK 
     & $\begin{array}{l} 
         D=4,5\ \text{OK} \\ D\ge6\ \text{?} 
       \end{array}$
     \\
\cline{2-6}
     &$\lambda>0$ & OK & OK  
     & $\begin{array}{l} 
         D\le6\ \text{OK} \\ D\ge7\ \text{?} 
        \end{array}$
     & $\begin{array}{l}
         D=4,5\ \text{OK} \\ D\ge6\ \text{?} 
        \end{array}$
     \\
\cline{2-6}
     &$\lambda<0$ & OK & OK  
     &  $\begin{array}{l}
          D=4\ \text{OK} \\ D\ge5\ \text{?} 
         \end{array}$
     &  $\begin{array}{l}
          D=4\ \text{OK} \\ D\ge5\ \text{?} 
         \end{array}$
     \\
\hline
$K=0$ &$\lambda<0$ & OK & OK  
     & $\begin{array}{l}
         D=4\ \text{OK} \\ D\ge5\ \text{?} 
        \end{array}$
     & $\begin{array}{l}
         D=4\ \text{OK} \\ D\ge5\ \text{?} 
       \end{array}$ 
     \\
\hline
$K=-1$ &$\lambda<0$ & OK & OK 
     & $\begin{array}{l}
         D=4\ \text{OK} \\ D\ge5\ \text{?} 
        \end{array}$
     & $\begin{array}{l}
         D=4\ \text{OK} \\ D\ge5\ \text{?} 
        \end{array}$
     \\
\hline
\end{tabular}
\end{center}
\end{table}

By applying the $S$-deformation to $V_{+}$ with
\Eq{
S=\frac{f}{h_+}\frac{dh_+}{dr};\ h_+=r^{n/2-1}H_+,
}
we obtain 
\Eq{
\tilde V_{S+}=\frac{k^2 f }{2r^2 H_+}
  \left[ (n-2)(n+1)\delta x + 2\right].
}
Since this is positive definite, the electromagnetic mode $\Phi_+$ 
is always stable for any values of $K$, $M$, $Q$ and $\lambda$, 
provided that the spacetime contains a regular black hole, although 
$V_{+}$ has a negative region near the horizon when $\lambda<0$ and 
$Q^2/M^2$ is small. 

Using a similar transformation, we can also prove the stability of 
the gravitational mode $\Phi_-$ for some special cases. For example, 
the $S$-deformation of $V_{-}$ with
\Eq{
S=\frac{f}{h_-}\frac{dh_-}{dr};\ h_-=r^{n/2-1}H_-
}
leads to
\Eq{
\tilde V_{-}=\frac{k^2f}{2r^2H_-}
  \left[ 2m-(n+1)(n-2)(1+m\delta)x \right].
}
For $n=2$, this is positive definite for $m>0$. When $K=1$, 
$\lambda\ge0$ and $n=3$ or when $\lambda\ge0,Q=0$ and the horizon is 
$S^4$, from $m\ge n+2$ ($l\ge2$) and the behaviour of the horizon value of $x$ (see Ref.\cite{Kodama.H&Ishibashi2004} for details), we can show that $\tilde V_{S-}>0$. Hence, in these special cases, the black hole is stable with respect to any type of perturbation. 

However, for the other cases, $\tilde V_{S-}$ is not positive 
definite for generic values of the parameters. The $S$-deformation 
used to prove the stability of neutral black holes is 
not effective either. Recently, Konoplya and Zhidenko studied the stability of this system for $n>2$ numerically. They found that if $\lambda\ge0$, the system is stable for $n\le 9$, i.e., $D\ge 11$\cite{Konoplya.R&Zhidenko2007}.

\subsection{Summary of the Stability Analysis}

The results of the stability analysis in this section can be summarised in Table \ref{kod:tbl:stability}. In this table, $D$ represents the spacetime dimension, $n+2$. The results for tensor perturbations apply only for maximally symmetric black holes, while those for vector and scalar perturbations are valid for black holes with generic Einstein horizons, except in the case with $K=1,Q=0,\lambda>0$ and $D=6$.

Note that this is a summary of the analytic study. As we mentioned above, the stability of AF/dS black hole is shown for $D<12$ numerically. Stability is shown also for topological adS black holes with non-positive mass\cite{Birmingham.D&Mokhtari2007A}.

\section{Flat black brane}
\label{kod:sec:FBB}

Static flat black brane solutions are perturbatively unstable in contrast to asymptotically simple static black holes discussed in the previous section. This was first shown by Gregory and Laflamme for the s-mode perturbation, i.e. perturbations that is spherically symmetric in the directions perpendicular to the brane\cite{Gregory.R&Laflamme1993,Gregory.R&Laflamme1995}. Later on, it was shown that the system has no other unstable modes numerically\cite{Seahra.S&Clarkson&Maartens2005,Kudoh.H2006}. These analyses however assumed that the frequency of an unstable mode, if it exists, is pure imaginary. In the static system this assumption may appear to be natural, but it is not the case in reality. In this section, we explain this point explicitly by applying the the gauge-invariant formulation in the previous section to this system. 

\subsection{Strategy}

Let us rewrite the $(m+n+2)$-dimensional flat black brane solution
\Eq{
ds^2=  -f(r) dt^2 + f(r)^{-1} dr^2 +r^2 d\sigma_n^2+ d\bm{x}^2,
}
which is the product of $(n+2)$-dimensional static black hole solution and the $m$-dimensional Euclidean space, as
\Eq{
ds^2 = g_{ab}(y) dy^a dy^b + r^2 d\sigma_n^2
}
with the $(m+2)$-dimensional metric
\Eq{
ds_{m+2}^2= g_{ab}(y) dy^a dy^b 
  =-f(r) dt^2 + f(r)^{-1} dr^2 + d\bm{x}^2.
}
Then, we can classify metric perturbations into tensor, vector and scalar types with respect to the $n$-dimensional constant curvature space $\K^n$ with the metric $d\sigma_n^2=\gamma_{ij}(z)dz^i dz^j$, and apply the gauge-invariant formulation developed in the previous chapter to them.  Further, since the background spacetime is homogeneous in the brane direction $\bm{x}$, for each type of perturbations, we can apply the Fourier transformation with respect to $\bm{x}=(x^p)$ to the perturbation variable as
\Eq{
\delta g_{\mu\nu}= h_{\mu\nu}(t,r,z^i) e^{ik\cdot x}.
}
Since the background metric is static, we can further apply the Fourier transformation with respect to $t$ to $h_{\mu\nu}$ if necessary and assume that
\Eq{
h_{\mu\nu}\propto e^{-i\omega t}.
}
Hence, we can reduce the Einstein equations for perturbations to a set of ODEs with respect to $r$. In this section, we assume that $\K^n$ is compact.

\subsection{Tensor perturbations}

The equation for tensor perturbations \eqref{kod:BulkPerturbationEq:tensor} with $\tau_T=0$ reads for the present system 
\Eq{
-\partial_t^2 H_T + \frac{f}{r^n}\partial_r(r^n f\partial_r H_T)
  -f\inpare{\frac{k_T^2+2K}{r^2}+k^2}H_T=0.
}
Let us define the energy integral for a tensor perturbation by 
\Eq{
E:=\int_{r_h}^\infty dr \, r^n
 \insbra{\frac{1}{f}\dot H_T^2 + f(H_T')^2
     +\inpare{\frac{k_T^2+2K}{r^2}+k^2}H_T^2}.
}
Then, from the perturbation equation, we have
\Eq{
\dot E=2\insbra{r^n f \dot H_T H_T'}_{r_h}^\infty
}
If there exists an unstable solution $H_T\propto e^{-i\omega t}$ with $\Im\omega<0$, it must fall off exponentially at $r\tend\infty$ and vanish at the horizon from the above equation, provided that the solution is uniformly bounded. For such a solution, $E$ becomes constant and contradicts the assumed exponential growth because all terms in the energy integral is non-negative definite. Hence, the black brane solution is stable for tensor perturbations.

\subsection{Vector perturbations}

\subsubsection{Basic perturbation equations}

Basic gauge-invariant variables for vector perturbations are given by $F^a(t,r)$ with $a=t,r,p$ ($p=1,\cdots,m$). Among these components, we decompose the part parallel to the brane,  $F_p$, into the longitudinal component $F_k$ proportional to the wave vector $k^p$ and the transversal components $F_p^\orth$ as
\Eqrsub{
&& F_k=ik^p F_p =\partial_p F^p,\\
&& F_p^\orth=F_p + \frac{ik_p}{k^2}F_k.
}

With this decomposition, the perturbation equations \eqref{kod:BulkPerturbationEq:vector1} and \eqref{kod:BulkPerturbationEq:vector1} can be written as
the four wave equations
\Eqrsub{
\hspace*{-1cm}&& \frac{1}{f}\partial_t^2F_t 
   -\frac{f}{r^n}\partial_r(r^n \partial_r F_t)
   +\frac{nf+ m_v + r^2k^2}{r^2}F_t
   =\inpare{f'-\frac{2f}{r}}\partial_t F_r,\\
\hspace*{-1cm}&& \frac{1}{f}\partial_t^2 F^r
   -\frac{f}{r^{n-2}}\partial_r(r^{n-2}\partial_r F^r)
   +\frac{2(n-1)f+m_v+k^2r^2}{r^2}F^r
   =\frac{f'}{f}\partial_t F_t,\\
\hspace*{-1cm}&& \frac{1}{f}\partial_t^2 F_k
   -\frac{1}{r^n}\partial_r(r^nf \partial_r F_k)
   +\frac{rf'+nf+m_v+k^2r^2}{r^2}F_k
   =\frac{2k^2 }{r}F^r,\\
\hspace*{-1cm}&& \frac{1}{f}\partial_t^2 (F^\orth/r)
   -\frac{1}{r^{n+2}}\partial_r\insbra{r^{n+2}f\partial_r(F^\orth/r)}
     +\inpare{k^2+\frac{m_v}{r^2}}(F^\orth/r)=0,
}
and the constraint
\Eq{
 -\frac{1}{f}\partial_t F_t 
   +\frac{1}{r^{n-1}}\partial_r(r^{n-1}f F_r)
   +F_k=0.
}
With the help of this constraint, the second of the above can be also written as
\Eq{
\frac{1}{f}\partial_t^2 F^r
-\frac{1}{r^{n-2}}\partial_r\inpare{r^{n-2}f\partial_r F^r}
  +\frac{(n-1)(2f-rf')+m_v + k^2r^2}{r^2}F^r
  =f' F_k.
}

Clearly, the transversal part $F^\orth_p$ decouple from the other modes and each component obeys the same single wave equation. Further, each of $(F_t,F^r)$ and $(F^r,F_k)$ obeys a closed set of equations, and the remaining components $F_k$ and $F_t$, respectively, are directly determined from them with the help of the above constraint equation. 

\subsubsection{Master equation}

Let us take $F^r$ and $F_k$ as fundamental variables and set
\Eq{
\Psi:=\begin{pmatrix} r^{n/2} F_k \\ (n+1)r^{n/2-1} F^r + r^{n/2}F_k \end{pmatrix}.
}
Then, the perturbation equations can be put into the form
\Eq{
\omega^2\Psi= \inpare{-D^2+V + fA}\Psi,
\label{kod:SMP:vector:basiseq}
}
where $V$ is the scalar potential
\Eq{
V=f\insbra{\frac{m_v}{r^2}+ k^2 + \frac{n(n+2)}{4r^2}f},
}
and $A$ is the matrix potential 
\Eq{
A=\begin{pmatrix} 
    \frac{2k^2}{n+1}+\frac{(n+2)f'}{2r} & -\frac{2k^2}{n+1} \\
    \frac{2k^2}{n+1} & -\frac{2k^2}{n+1}-\frac{n}{2r}f'
    \end{pmatrix}.
}

In order to see whether this set of equations can be reduced into decoupled single equations, we introduce a new vector variable $\Phi$ by
\Eq{
\Phi= Q \Psi + P \Psi',
}
where $P$ and $Q$ are matrix functions of $r$ that are independent of $\omega$. If we require that $\Phi$ obeys the equation of the form
\Eq{
\Phi'' + (\omega^2-V-W)\Phi=0
}
with a diagonal matrix $W$, we obtain constraints on $V$ and $B$. 

For the exceptional mode with $m_v=0$, these constraints are satisfied, and we find that  for the choice $P=1$ and 
\Eq{
Q=\begin{pmatrix} 
       -\frac{k^2 r}{n+1}-\frac{n+2}{2r}f & \frac{k^2r}{n+1} \\
       -\frac{k^2 r}{n+1} & \frac{k^2 r}{n+1}+\frac{n}{2r}f
            \end{pmatrix},
}
$W$ is given by the diagonal matrix whose entries are
\Eq{
W_1=\frac{n+2}{r^2}f\inpare{1-\frac{(n+1)M}{r^{n-1}}},\quad
W_2=-\frac{n}{r^2}f\inpare{1-\frac{(n+1)M}{r^{n-1}}}.
}
The corresponding equations for $\Phi$ decouple to
\Eqr{
&& \Phi_i''+(\omega^2-V_i)\Phi_i=0,\\
&& V_1=f\insbra{k^2+\frac{n+2}{4r^2}
              \inpare{n+4-\frac{2(3n+2)M}{r^{n-1}}}},\\
&& V_2=f\insbra{k^2+\frac{n}{4r^2}
              \inpare{n-2+\frac{2nM}{r^{n-1}}}}.
}
$V_2$ is clearly positive. Further, in terms of the S-deformation with
\Eq{
S=\frac{n+2}{2r}f
}
$V_1$ is transformed into 
\Eq{
\tilde V_1=k^2 f>0
}
Hence, this system is stable for this exceptional mode.

If we apply the same transformation in the case $m_v\neq0$, we obtain
\Eq{
\insbra{(f\pd_r)^2
-\frac{2m_v fh}{r(r^2\omega^2-m_v f)}f\pd_r+\omega^2-V_0}\Phi
=\frac{fh}{(n+1)(r^2\omega^2-m_v f)} B\Phi,
}
where
\Eqr{
&& h=1-\frac{(n+1)M}{r^{n-1}},\\
&& V_0=f\insbra{\frac{m_v}{r^2}+k^2+\frac{n^2+2n+4}{4r^2}
       -\frac{(n^2+4n+2)M}{2r^{n+1}}
       +\frac{m_vfh}{r^2(r^2\omega^2-m_v f)}},\notag\\
&&\\
&& B=\begin{pmatrix} 
     (n+1)^2\omega^2+2m_v k^2 & -2m_v k^2\\
     2m_v k^2  & -\inrbra{(n+1)^2\omega^2+2m_v k^2}
     \end{pmatrix}
}
Since $B$ is a constant matrix with eigenvalues
\Eq{
\lambda=\pm (n+1)\omega \insbra{(n+1)^2\omega^2+4m_v k^2}^{1/2},
}
we can reduce the set of equations for $\Phi$ to decoupled single second-order ODEs. However, these equations are not useful in the stability analysis because their coefficients depend on $\omega^2$ nonlinearly and have singularities in general\footnote{In Ref\cite{Kudoh.H2006} the author derived a well-behaved single master equation of 2nd-order for the black string background. There, the author took the gauge in which  $f_z=0$ and $H_T=0$. Such a gauge cannot be realised in general because the gauge transformations of $f_z$ and $H_T$ are given by $\b\delta f_z=-S \pd_z(L/S)$ and $\b\delta H_T= k_v L/S$. If we set $f_z=0$, we cannot change the $z$-dependence of $H_T$ in general.}.

\subsubsection{Stability analysis}

Since we cannot find a convenient master equation, let us try to analyse the stability by directly looking into the structure of the set of equations \eqref{kod:SMP:vector:basiseq}. The subtle point of this set of equations is that the operator on the right-hand side is not self-adjoint because $A$ is not a hermitian matrix. Therefore, we cannot directly conclude that $\omega^2$ is real. 

Allowing for the possible existence of the imaginary part of $\omega^2$, we obtain the following two integral relations from the above equation:
\Eqrsub{
\hspace*{-1cm}&& \Re(\omega^2)(\Psi,\Psi)
   =\int_{r_h}^\infty\frac{dr}{f}
   \insbra{ (D\Psi_1)^2+(D\Psi_2)^2
    +f U_1 |\Psi_1|^2   +f U_2 |\Psi_2|^2 },\\
\hspace*{-1cm}&& \Im(\omega^2)(\Psi,\Psi)=-\frac{4k^2}{n+1}
   \int_{r_h}^\infty dr \Im(\bar\Psi_1 \Psi_2).
}
Here, $D=f d/dr$ and 
\Eqrsub{
&& U_1=\frac{m_v}{r^2}+\frac{n+3}{n+1}k^2
    +\frac{n(n+2)}{4r^2}f +\frac{(n+2)f'}{2r},\\
&& U_2=U_1-\frac{4k^2}{n+1}-\frac{n+1}{r}\frac{f'}.
}
By applying the S-deformation with 
${
S=\frac{n }{2r}f
}$
to $\Psi_2$, the right-hand side of the equation corresponding to $\Re(\omega^2)$ is deformed to
\Eq{
D\Psi_2 \maps (D+S)\Psi_2,\quad
U_2 \maps \frac{m_v}{r^2}+\frac{n-1}{n+1}k^2.
}

Therefore, if we assume that $\omega^2$ is real, as is assumed in most work, we can conclude that the system is stable against vector perturbations.  However, we cannot exclude the possible existence of an unstable mode with $\Im(\omega^2)\neq0$. 

\subsection{Scalar perturbations}

\subsubsection{Perturbation variables}

The gauge-invariant variable set $F_{ab}$ in the general formulation can be decomposed into the scalar, vector and tensor parts by their transformation behavior with respect to the brane coordinates as
\Bitm
\item[] Scalar part: $F_{tt}, F_{tr}, F_{rr}, F_{kt}, F_{kr}, F_{kk},
   F_\orth$.
\item[] Vector part: $F_{\orth p t}, F_{\orth p r}, F_{\orth p k}$.
\item[] Tensor part: $F_{\orth p\,\orth q}$.
\Eitm
Here,
\Eqrsub{
&& F_{k a}=\partial^p F_{p a}/(ik)=(k^p/k) F_{p a},\\
&& F_{\orth p a}=F_{p a}-(k_p/k^2) k^q F_{q a}
                =F_{p a}-(k_p/k) F_{k a},\\
&& F_{kk}= (k^p k^q /k^2) F_{pq},\\
&& F_\orth= F^p_p - F_{kk},\\
&& F_{\orth p\,\orth q}= F_{\orth p q}-(k_q/k) F_{\orth p k}
     -\frac{1}{d-1} F_\orth (\delta_{pq}- k_p k_q/k^2).
}
The remaining gauge-invariant variable $F$ in the general formulation also belongs to the scalar part. 
Note that the vector and tensor parts do not exist for the black string background.

\subsubsection{S-mode}

First, we consider the exceptional mode with $k_s=0$, which is often called the S-mode. For this exceptional mode, the general gauge-invariant variables reduce to $F_{ab}=f_{ab}$ and $F=H_L$ due to the non-existence of corresponding harmonic vectors and tensors. These variables are not gauge invariant and subject to the gauge transformation law
\Eqrsub{
&& \bar\delta H_L= - \frac{f}{r} T_r,\\
&& \bar\delta f_{tt}=2i\omega T_t + ff' T_r,\quad
   \bar\delta f_{tr}=i\omega T_r - f(T_t /f)',\notag\\
&& \bar\delta f_{rr}=-2T_r'-(f'/f) T_r,\\
&& \bar\delta f_{tk}=i\omega T_k -i k T_t,\quad
   \bar\delta f_{rk}=-T_k' - i k T_r,\quad
   \bar\delta f_{kk}= -2i k T_k,\\
&& \bar\delta f_{\orth p t}=i\omega T_{\orth p},\quad
   \bar\delta f_{\orth p r}=-T_{\orth p}',\quad
   \bar\delta f_{\orth p k}=-i k T_{\orth p},\\
&& \bar\delta f_{\orth}=0,\quad
   \bar\delta f_{\orth p\,\orth q}=0.
}
In particular, we have
\Eq{
\bar\delta\inpare{f_{tk}+\frac{\omega}{2k}f_{kk}}=-i k T_t,\quad
\bar\delta\inpare{f_{rk}+\frac{i}{2k}f_{kk}'}=-i k T_r.
}

From these, we can construct the following five gauge invariants for the scalar part:
\Eqrsub{
&& r^{2-n}X=-F_\orth,\\
&& r^{2-n}(X-Y)=F^r_r-2rF'-\inpare{\frac{rf'}{f}-2}F,\\
&& r^{2-n}Z=F^r_t+\frac{if^2}{2\omega}(F^t_t)'+i\omega r F
            -\frac{if^2}{2\omega}\inpare{\frac{rf'}{f}F}',\\
&& r^{2-n} V^t= F^t_k + \frac{k}{2\omega}\inpare{F^t_t-\frac{rf'}{f}F}
              -\frac{\omega}{2kf}F_{kk},\\
&& r^{2-n}V^r=F^r_k -ikr F + \frac{if}{2k}F_{kk}'
}
For the vector part, we adopt the following two gauge invariants
\Eqrsub{
&& r^{2-n} W^t_p = F^t_{\orth p}-\frac{\omega}{kf}F_{\orth p k},\\
&& r^{2-n} W^r_p = F^r_{\orth p}+ \frac{if}{k} F_{\orth p k}'.
}
%

\paragraph{(1) Tensor part.}

First, we study the stability in the tensor part. The perturbation variable of this part, $F_{\orth p\,\orth q}$, follows the closed equation
\Eq{
-f(r^n f F_{\orth p\orth q}')'
       +(k^2 f-\omega^2)r^n F_{\orth p\orth q}=0.
\label{kod:BB:SP:Swave:orthorth:BasicEq}
}
From this we obtain the integral relation
\Eq{
\omega^2 \int_{r_h}^\infty dr_* r^n |F_{\orth p\orth q}|^2
 =\int_{r_h}^\infty dr r^n \insbra{
   f|F_{\orth p\orth q}'|^2
   +k^2 |F_{\orth p\orth q}|^2}
   -\insbra{r^n f \bar F^{\orth p\orth q}
       F_{\orth p\orth q}'}_{r_h}^\infty.
}

If there exists an unstable mode with 
\Eq{
\omega=\omega_1+ i\omega_2;\quad \omega_2>0.
}
a solution that is bounded at the horizon behaves as
\Eq{
F_{\orth p\,\orth q}\sim e^{ -i\omega r_*}
}
near the horizon. Next, at infinity, the solution behaves
\Eq{
F_{\orth p\,\orth q}\sim \frac{1}{r^{(n-1)/2}}Z_\nu (\sqrt{\omega^2-k^2}r)
 \sim r^{-n/2}\exp (\pm i\sqrt{\omega^2-k^2}r).
}
Therefore, for an unstable mode that is uniformly bounded, the boundary term in the above integral relation vanishes and the integral at the left-hand side converges. This implies that $\omega^2>0$ and leads to contradiction. 

\paragraph{(2) Vector part.}
Next, for the vector part, we obtain the following two equations for the gauge-invariant variables $W^t_p$ and $W^r_p$:
\Eqrsub{
&& -i\omega\insbra{W^t_{\orth p}{}'
         -\inpare{\frac{n-2}{r}+\frac{f'}{f}}W^t_{\orth p}}
  +\frac{k^2 f-\omega^2}{f^2}W^r_{ \orth p}=0,\\
&&  -i\omega W^t_{\orth p}+W^r_{\orth p}{}' 
       +\frac{2}{r}W^r_{\orth p}=0.
}
Therefore, we can set 
\Eq{
W^r_{\orth p}= r^{-2}\Phi,\quad 
i\omega W^t_{\orth p}=r^{-2}\Phi',
}
and the perturbation equations can be reduced to the following single master equation for $\Phi$;
\Eq{
-f(r^{-n}f \Phi')' + (k^2 f-\omega^2) r^{-n}\Phi=0.
}
By the same argument for the tensor part, we can show that this equation does not have a uniformly bounded solution with $\Im(\omega)>0$.

\paragraph{(3) Scalar part.}

Finally for the scalar part, the perturbation equations gives the closed 1st-order set of equations for $X,Y,Z, V^t$,
\Eqrsub{
& X' =
 &\frac{1}{k^2r Hf^2}\left[ r^2\omega^4
   -\omega^2\inrbra{k^2r^2f+n-n(n+1)x+\frac{3n^2+2n-1}{4}x^2}
   \right.\notag\\
&& \left.  -\inpare{2+\frac{n-5}{2}x}k^2Hf\right] X
    +\frac{1}{k^2rfH}\inrbra{n\omega^2\inpare{1-\frac{n+1}{2}x}
    +k^2H^2}Y  \notag\\
&&    +\frac{2i\omega}{k^2f^2H}(n\omega^2-k^2H)Z
    +\frac{\omega}{krfH}\inrbra{2\omega^2r^2
          +(n-1)xH}V^t, \\
& Y' =
 & \frac{1}{k^2rf^2H}\left[r^2\omega^4
    -\omega^2\inrbra{2k^2r^2f+n-n(n+1)x+\frac{3n^2+2n-1}{4}x^2}
    \right.\notag\\
&& \qquad \left.
  +r^2k^4f^2 -\inrbra{n-(n^2+1)x+\frac{(n+1)^2}{4}x^2}k^2f
     \right]X  \notag\\
&& +\frac{1}{k^2rfH}\insbra{n\omega^2\inpare{1-\frac{n+1}{2}x}
    +(n-1)k^2\inrbra{n-\frac{5n}{2}x+\frac{3(n+1)}{4}x^2}}Y
   \notag\\
&& +\frac{2in\omega}{k^2f^2H}(\omega^2-k^2f)Z
  +\frac{\omega}{krfH}\inrbra{2r^2\omega^2-2k^2r^2f
   +(n-1)xH}V^t,\\
& Z'=
 & -i\frac{(n-1)^2x}{2\omega r^2}X
   +\frac{i}{2r^2\omega}\inrbra{r^2\omega^2+(n-1)^2x}Y
   \notag\\
&& -\frac{2}{rf}\inpare{1-\frac{n+1}{2}x}Z
   +ikf V^t,\\
& (V^t)' =
 & \frac{\omega}{2k^3f^3H}\left[ -r^2\omega^4
    +\omega^2\inrbra{2k^2r^2f +n-n(n+1)x+\frac{3n^2+2n-1}{4}x^2}
    \right.\notag\\
&&\qquad \left.
  -r^2k^4 f^2 -k^2 f\inrbra{n-2n^2 x+\frac{5n^2+2n-3}{4}x^2}
   \right] X \notag\\
&& +\frac{\omega}{2rk^3f^2H}\insbra{-n\omega^2\inpare{1-\frac{n+1}{2}x}
    +k^2\inrbra{n-n(n+1)x+\frac{3n^2+2n-1}{4}x^2}}Y
    \notag\\
&& +\frac{i}{k^3f^3H}\inrbra{-n\omega^4
   +k^2\omega^2\inpare{2n-\frac{3n+1}{2}x}-k^4fH}Z
   \notag\\
&& +\frac{1}{rk^2f^2H}\inrbra{-r^2\omega^4
   +\omega^2\inpare{k^2r^2f-\frac{n-1}{2}xH}
   +(n-2)k^2f^2H}V^t.
}
and the expression for $V^r$ in terms of these quantities,
\Eq{
V^r = \frac{i(\omega^2+k^2f)}{2nk^3 rf^2}
  \insbra{ \inrbra{(\omega^2+k^2f)r^2+nf^2} X
  +nf^2 Y + \frac{2in\omega^3 r}{\omega^2+k^2f} Z
  +2\omega k r^2 f V^t }.
}
Here, $x=2M/r^{n-1}$.

From these equations, we find that $X$ obeys the closed 2nd-order ODE
\Eqr{
&& -f(fX')'+(n-4)\frac{f^2}{r}X' \notag\\
&&\qquad +\insbra{-\omega^2
    +f\inpare{k^2 + \frac{n-2}{r^2}\inrbra{1+(n-2)x}}}X=0,
}
which can be put into the canonical form in terms of $\Phi$ defined by
\Eq{
X=r^{n/2-2}\Phi
}
as 
\Eq{
-f(f\Phi')'+\insbra{-\omega^2
    +f\inrbra{k^2+\frac{n}{4r^2}(n-2+nx)}}\Phi=0.
}
It is clear that this equation does not have an unstable mode.

Next, let us define the new variable $\Omega$ by
\Eqr{
&& \Omega:=P X + nf\inpare{1-\frac{n+1}{2}x}Y + 2in\omega r Z
       +2k\omega r^2 f V^t ;\\
&& P:= \insbra{\frac{n+1}{2n}x
     -\frac{(n-1)x}{2k^2r^2}\inpare{n-\frac{n+1}{2}x}}\omega^2 r^2
     +\frac{n-1}{2n}xk^2 r^2 \notag\\
&& \qquad\qquad 
  -n+n(n+1)x -(3n^2+2n-1)x^2,
}
Then, we find that $\Omega$ satisfies a closed 2nd-order ODE $\mod\ X=0$:
\Eq{
-f(f\Omega')'+ A f\Omega' +(-\omega^2+V_\Omega)\Omega=B X,
}
where
\Eqrsub{
&& A=\frac{f}{4r^3gH}\left[\inrbra{4n^2+2(n+1)(n-2)x}k^2r^2
   \right. \notag\\
&&\qquad\qquad\left.
      +n^2(n-1)x\inrbra{3(n+1)x-2(n+2)}\right],\\
&& V_\Omega=\frac{f}{8r^4 g^2H}\left[
    2\inrbra{2n-(n+1)x}^2 k^4r^4 \right.\notag\\
&&\qquad\qquad    
  +\left\{8n^2(n+2)-4n(n+2)(3n^2+n+2)x \right.\notag\\
&&\qquad\qquad \left.
   +2n(n+1)(8n^2+5n+5)x^2 -(n+2)(3n-1)(n+1)^2 x^3\right\}k^2 r^2
   \notag\\
&&\qquad\qquad
   +n^2(n-1)x\left\{n(n+1)^2x^3-3(3n-1)(n+1)x^2 \right.\notag\\
&&\qquad\qquad\left.\left.    +4(2n^2+2n-1)x-4n^2\right\} \right],\\
&& B=\frac{fg}{nr^2H}\inrbra{(n+1)\omega^2-k^2}
   \insbra{-2k^2r^2 (1-nx)+n(n-1)x(n-x)},\\
&& H:=k^2+\frac{n(n-1)}{2r^2}x,\quad
 g:=n-\frac{n+1}{2}x.
}

\begin{figure}[t]
\centerline{%
\includegraphics[width=6cm]{\FigDir/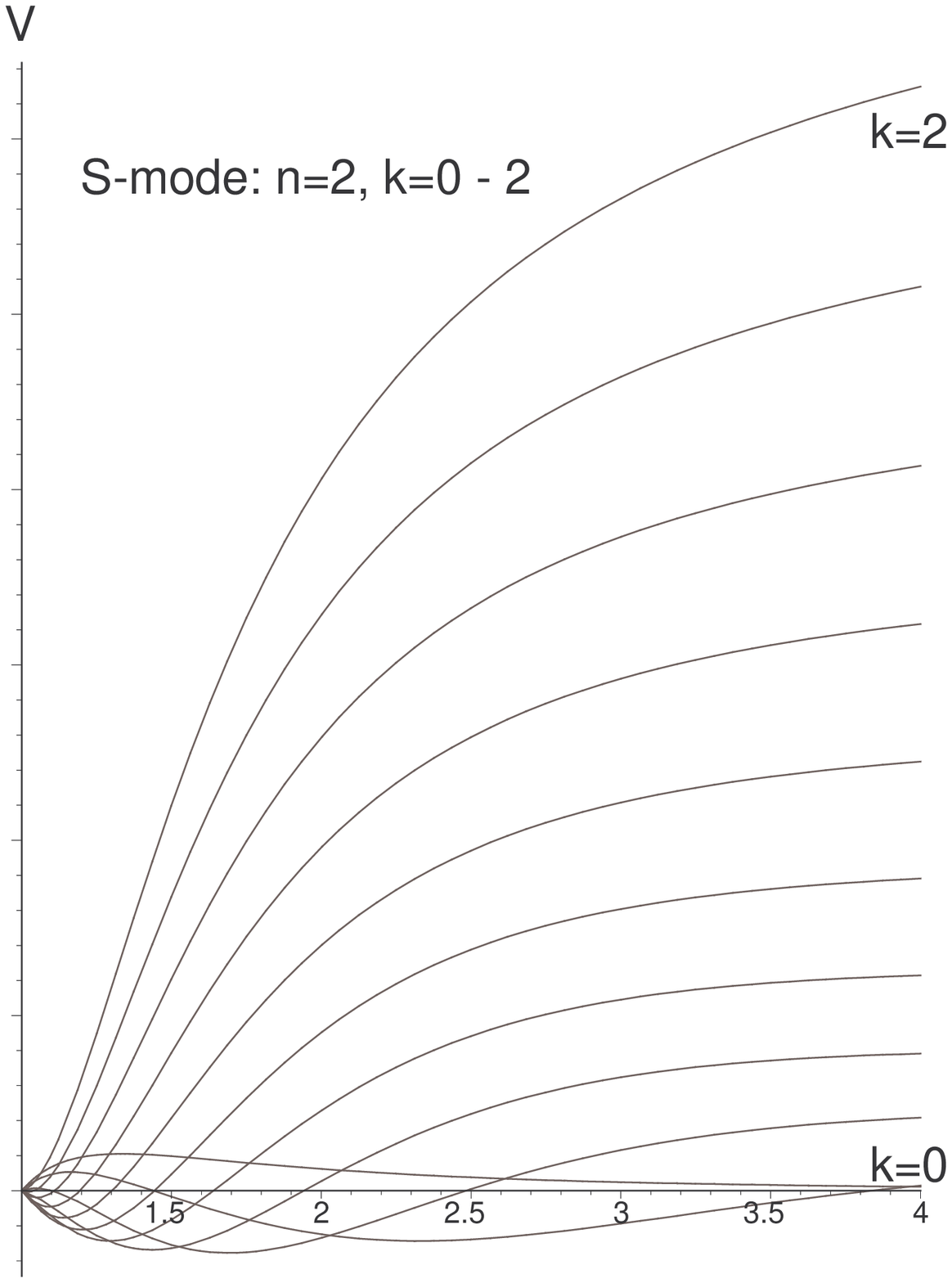}%
\hspace{1cm}%
\includegraphics[width=6cm]{\FigDir/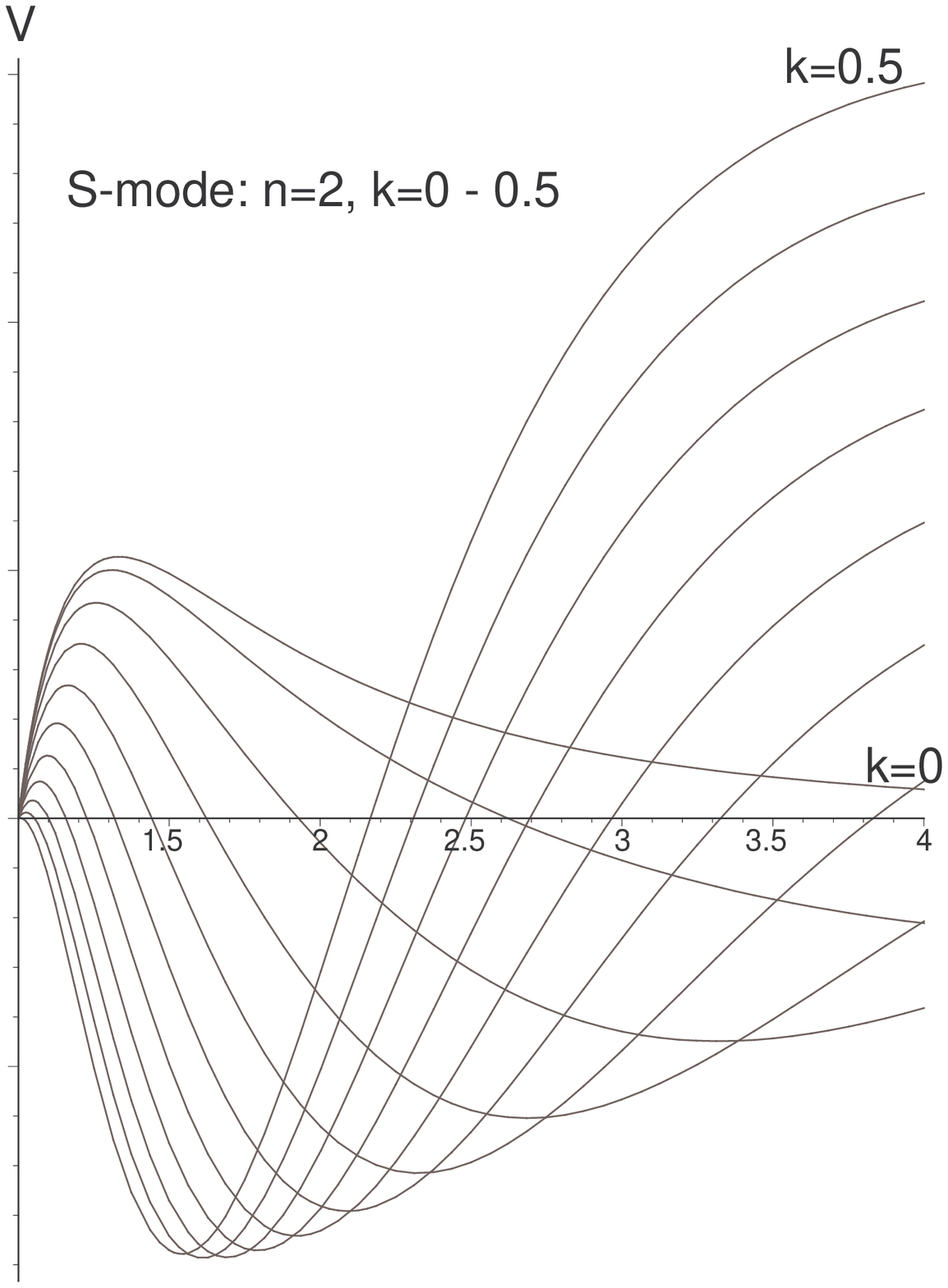}}
\caption{The effective potential for S-modes}
\label{kod:fig:PotSmode}
\end{figure}

By the transformation
\Eq{
 \Omega=r^{n/2}g H \Psi,\\
}
we can put this equation into the canonical form
\Eq{
-f(f\Psi')'+(-\omega^2+V)\Psi= r^{n/2}gH B X,
}
where
\Eqr{
&V=& \frac{f}{H^2}U;\\
&U=& k^6 + \frac{(n+4)k^4}{4r^2}(n+2-3nx) \notag\\
&& -\frac{n(n-1)k^2}{4r^4}\inrbra{3n(n+2)-(2n^2+3n+4)x}x
  \notag\\
&& +\frac{n^3(n-1)^2}{16r^6}x^2(n-2+nx).
}
This potential has a deep negative region for $0<k<k_n$ with some constant $k_n$ dependent on $n$. It has been shown by numerical calculations\cite{Seahra.S&Clarkson&Maartens2005,Kudoh.H2006} that the eigenvalue $\omega^2$ becomes negative for some range $0<k<k_c$, as first pointed out by Gregory and Laflamme using a different reduction\cite{Gregory.R&Laflamme1993}.

\subsubsection{Generic scalar perturbation}

\paragraph{(1) Tensor part.} 
The tensor part of generic scalar-type perturbations obeys the decoupled 2nd-order ODE

\Eq{
-f(r^n f F_{\orth p\,\orth q}')' 
   +\inpare{-\omega^2+k^2 f+\frac{n+m}{r^2}f}r^nF_{\orth p\,\orth q} =0.
}
It is obvious that this equation has no unstable mode.

\paragraph{(2) Vector part.}

In terms of the gauge-invariant fundamental variables
\Eq{
V_t= r^{n-2} F_{t\orth p},\quad
V_r= r^{n-2} F_{r\orth p},\quad
V_k= r^{n-2} F_{k \orth p}, 
}
the perturbation equations for the vector part are expressed as
\Eqrsub{
& &-(fV_r)'-i\omega f^{-1} V_t-ik V_k =0,\\
&  & i\frac{\omega}{f}\inpare{ V_t' -\frac{n-2}{r}f V_t}
        +\inpare{-\frac{\omega^2}{f}+k^2+\frac{n+m}{r^2}}V_r
 +ik \inpare{V_k' -\frac{n-2}{r}V_k} =0,\\
&& -r^{n-4}(r^{4-n}f V_k')'
   +\inpare{-\frac{\omega^2}{f}+\frac{n+m}{r^2}
      +\frac{n-2}{r}f'+\frac{n-2}{r^2}f}V_k \notag\\
&& \qquad
 -\frac{k\omega}{f} V_t + ik \inpare{(fV_r)'+\frac{2f}{r}V_r}=0, \\
&& -r^{n-4}f (r^{4-n} V_t')'
     +\inpare{\frac{m+n}{r^2}+\frac{n-2}{r^2}}V_t
  -i\omega f\inpare{V_r'+\frac{2}{r}V_r} +k\omega V_k=0.
}

By eliminating $V_t$ and introducing the new variables $Y$ and $Z$ by
\Eq{
\Phi=\begin{pmatrix} Z \\ Y \end{pmatrix};\quad
   V_k= -i r^{n/2-2}Z,\quad 
   V_r= f^{-1} r^{n/2-1}Y,
}
this set of equations are reduced to a set of two ODEs,
\Eqr{
&& D^2 \Phi - \inrbra{-\omega^2+k^2 f+\frac{n+m}{r^2}f
  +\frac{n(n-2)}{4}f^2}\Phi
  = A \Phi; \\
&& A =\begin{pmatrix} 
      \frac{nf'}{2r}f  &  -2k f \\
      -\frac{kf'}{r}f  &   -\frac{(n-2)f'}{2r}f
      \end{pmatrix}.
}
This set of equations has the same structure as that for vector perturbations and can be shown to have no unstable mode if $\omega^2$ is real.

\paragraph{(3) Scalar part.}

Finally, we discuss the scalar part of the generic scalar-type perturbation. Utilising one of the Einstein equations
\Eq{
E_T\equiv 2(n-2)F+F^a_a=0,
}
the basic perturbation variables can be expressed in terms of $X,Y,Z,V^t,V^r,S$ and $\Psi$ as
\Eqrsub{
&& \tilde F^t_t=X + 2\tilde F -k^2 f V^t,\quad
   \tilde F^r_r=Y + 2\tilde F,\quad
   \tilde F^r_t=i\omega Z,\\
&& \tilde F^r_k=ik V^r,\quad
   \tilde F^t_k=\omega k V^t,\quad
   \tilde F_{kk}=S +\omega^2 V^t +2\tilde F,\\
&& 2(n+1)\tilde F=-\Psi-X-Y-S-(\omega^2-k^2 f)V^t,\quad
   \tilde F_\orth=\Psi.
}
Here, $\t Q=r^{n-2} Q$ in general.

In terms of these variables, the Einstein equations can be reduced to the decoupled single equation for $\Psi$, 
\Eq{
-r^{-n}f(r^n f\Psi')'
   +\insbra{-\omega^2+\inpare{k^2+\frac{n+m}{r^2}}f}\Psi=0.
}
and the regular 1st-order set of ODEs for $X,Y,Z,V^t,V^r$ and $S$,
\Eqrsub{
&& Z' = X,\\
&& X' = \frac{n-2}{r}X+\inpare{\frac{f'}{f}-\frac{2}{r}}Y
  +\frac{1}{f}\inpare{-\frac{\omega^2}{f}+k^2+\frac{m+n}{r^2}}Z
  \notag\\
&&\qquad\qquad
   +k^2 f' V^t,\\
&& Y' = \frac{f'}{2f}(X-Y)+\frac{\omega^2}{f^2}Z
  +\frac{k^2}{f}\inpare{V^r-\frac{ff'}{2}V^t},\\
&& (V^r)' = -S,\\
&& S' = \frac{n-2}{r}S-\frac{2}{r}Y + \omega^2\frac{f'}{f}V^t
  +\frac{1}{f}\inpare{\frac{\omega^2}{f}-k^2-\frac{n+m}{r^2}}V^r,\\
&& k^2r^2 f'f^2 (V^t)' =
   \insbra{2\omega^2r^2+(n-1)x\inpare{n-\frac{n+1}{2}x}}X
   \notag\\
&&\qquad
   +\insbra{2\omega^2 r^2-2(k^2r^2+n+m)f+2n -4nx+\frac{(n+1)^2}{2}x^2}Y
  \notag\\
&&\qquad
   +\frac{1}{r}\insbra{-2n\omega^2 r^2+(n-1)x(k^2r^2+n+m)}Z
\notag\\
&&\qquad
   -(n-1)k^2 x\inpare{2+\frac{n-5}{2}x} fV^t
   -2k^2r\inpare{n-\frac{n+1}{2}x}V^r
\notag\\
&&\qquad -2k^2r^2 fS,
}
where $x=2M/r^{n-1}$.

If we define $X_1,X_2,X_3$ by
\Eqrsub{
&& X_1=Z,\\
&& X_2=r(X+Y)-nZ + k^2 rf V^t,\\
&& X_3=-\inpare{1+\frac{rf'}{2nf}}\insbra{r(X+Y)-nZ}
       -\frac{k^2r^2}{2n}f' V^t,
}
and introduce $\Phi$ by
\Eq{
\Phi:=\begin{pmatrix} r^{-n/2}X_1 \\ r^{-n/2+1}f^{-1}X_2 \\ r^{-n/2}X_3
      \end{pmatrix},
}
we can reduce the above set of 1st-order ODEs to the set of 2nd-order ODEs of the normal eigenvalue type as
\Eq{
\omega^2 \Phi = (-D^2+ V_0 + W )\Phi.
}
Here, $V_0$ is the scalar potential
\Eq{
V_0=\frac{f}{4r^2}\insbra{4(m+k^2r^2)+n^2-2n+n(n+4)x},
}
and $W$ is the following matrix of rank 3:
\Eqrsub{
&& W_{11}=0,\quad
   W_{12}=\frac{(n^2-1) xf^2}{nr^3},\quad
   W_{13}=\frac{2f^2}{r^2},\\
&& W_{21}=\frac{\inrbra{4-2(n+1)x}k^2r^2 -2(n-1)mx
         -n(n^2-1)x}{rf},\\
&& W_{22}=\frac{nf^2}{r^2},\quad
   W_{23}=0,\\
&& W_{31}=\frac{1}{2nr^2f}\left[
     2(n-1)x(n-2+x)r^2k^2+\inrbra{4n+2n(n-5)x+2(n+1)x^2}m
     \right.\notag\\
&&\qquad\quad\left.
   +n(n+1)x\inrbra{2n^2-(2n^2+3n-1)x+n(n+1)x^2}\right],\\
&& W_{32}=\frac{(n^2-1)x\inrbra{2-(n+1)x} f}{2nr^3},\quad
   W_{33}=\frac{(n+1)\inrbra{2-(n+1)x}f}{r^2}.
}

The original variables $X,Y,Z,Vt,Vr,S$ can be expressed in terms of $X_i$ and $DX_i$ as
\Eqrsub{
&& X=X_1',\quad
  Y=-X_1'+\frac{n}{r}X_1 -\frac{(n-1)x}{2nrf}X_2 -\frac{1}{r}X_3,\quad
   Z=X_1,\\
&& V^t=-\frac{(n+1) x-2n)}{2nk^2 rf^2} X_2+ \frac{1}{k^2 rf} X_3,\\
&& V^r= -\frac{(n-1)x}{2n k^2 r}X_2' -\frac{f}{k^2 r} X_3' 
 -\frac{2r^2k^2+2m+n(n+1)x}{2k^2r^2} X_1
\notag\\
&&\qquad
 +\frac{(n-1)x\inrbra{(n+1)x-2}}{2nk^2r^2f} X_2
 -\frac{f}{k^2r^2} X_3,\\
&& S=\frac{2k^2r^2+2m+n(n+1)x}{2k^2r^2 } X_1'
      +\frac{nf}{k^2r^2} X_3'
  +\frac{(n-1)x(2k^2r^2+2m+n+n^2)}{2k^2r^3f}X_1
  \notag\\
&&\qquad
  +\frac{-\omega^2r^2+f(n+m+k^2r^2)}{k^2r^3f}
   \insbra{ \frac{(n-1)x}{2nf}X_2 + X_3}.
}
Hence, the equation for $\Phi$ is equivalent to the original 1-st order system. 

Unfortunately, it is not possible to analyse the stability of this 2nd-order system by an analytic method, partly because it is not a self-adjoint system. However, all numerical calculations done by various authors have found no evidence of instability for this system\cite{Seahra.S&Clarkson&Maartens2005,Kudoh.H2006}.

\section{Summary and Discussion}
\label{kod:sec:Summary}

In this lecture, we have explained the gauge-invariant formulation for perturbations of a class of background solutions to the Einstein equations that include various practically important spacetimes as special cases. Then, we have illustrated its power by applying it to the stability problem of static black holes in higher dimensions and flat black branes. 

These two systems have one important common feature in addition to staticity that the background spacetime is of the cohomogeneity one. That is, the isotropy group of the spacetime has orbits with codimension one, and roughly speaking, the spacetime is inhomogeneous only in one direction, say $r$. In this case, the perturbation equations for the system can be automatically reduced to a set of ODEs for functions of $r$ with the help of the harmonic expansion. This applies to a rotating black hole case as well\cite{Murata.K&Soda2007A}

There exists however one crucial difference between the two systems. In the static black hole case, the perturbation equations can be reduced to decoupled single 2nd-order ODE, and the stability problem is formulated as an eigenvalue problem for the corresponding self-adjoint operator. In contrast, in the black brane case, it appears to be impossible to reduce all the perturbation equations to decoupled single master equations. Further, the eigenvalue problem for the stability issue cannot be put in the self-adjoint form even if we allow for a multi-component expression, except for some special modes. Nevertheless, numerical calculations indicate that there exists no eigen-mode with an imaginary frequency\cite{Seahra.S&Clarkson&Maartens2005,Kudoh.H2006}. There must be a profound reason behind this result.

The gauge-invariant formulation developed in \S\ref{kod:sec:GIPT} can be also applied to spacetimes whose cohomogeneity dimension is greater than one. Although this implies that the formulation can be applied to perturbations of rotating black holes in higher dimensions such as the Myers-Perry solution\cite{Myers.R&Perry1986} and its generalisation to non-vanishing cosmological constant\cite{Gibbons.G&&2005}, it may not be practically useful in most case, because we obtain a couple set of partial differential equations in a reduced spacetime with dimensions smaller than the original one. However, in some special cases, we obtain a single PDE that is separable to ODEs. For example, for a Kerr(-adS) black hole that rotates in a two-dimensional plane, we can classify perturbations into tensor, vector and scalar types as in the static case, and among these, the perturbation equation for the tensor-type perturbation is separable\cite{Kodama.H2007A}. There may exist other cases in which similar phenomena happen.

\section*{Acknowledgements}

The authour thanks the organisers of the 4th Aegean Summer School, especially Prof. Elefteris Papantopoulos for giving the author the chance to deliver a lecture at the beautiful island Lesvos. 
This work is partly supported by Grants-in-Aid for Scientific Research from JSPS (No. 18540265).



\end{document}